\documentclass[onecolumn,showpacs,aps,epsfig,nofootinbib]{revtex4}

\usepackage{graphicx}
\usepackage{epstopdf}
\usepackage{latexsym}
\usepackage{amssymb}
\usepackage{amsmath}
\usepackage{color}
\usepackage{datetime}
\usepackage{extarrows}
\usepackage{mathrsfs,amssymb,bm}
\usepackage{hyperref}
\usepackage{multirow}
\usepackage{makecell}

\usepackage[center]{subfigure}

\begin{document}

 \newcommand{\bq}{\begin{equation}}
 \newcommand{\eq}{\end{equation}}
 \newcommand{\bqn}{\begin{eqnarray}}
 \newcommand{\eqn}{\end{eqnarray}}
 \newcommand{\nb}{\nonumber}
 \newcommand{\lb}{\label}
 \newcommand{\tc}{\textcolor{black}}
 \newcommand{\beq}{\begin{equation}}
\newcommand{\eeq}{\end{equation}}
\newcommand{\be}{\begin{equation}}
\newcommand{\ee}{\end{equation}}
\newcommand{\PRL}{Phys. Rev. Lett.}
\newcommand{\PL}{Phys. Lett.}
\newcommand{\PR}{Phys. Rev.}
\newcommand{\CQG}{Class. Quantum Grav.}

\title{Preferred axis in cosmology}

\author{Wen Zhao, Larissa Santos}
 \affiliation{ CAS Key Laboratory for Researches in Galaxies and Cosmology, Department of Astronomy, University of Science and Technology of China, Chinese Academy of Sciences, Hefei, Anhui 230026, China}
%


\begin{abstract}

The foundation of modern cosmology relies on the so-called cosmological principle which states an homogeneous and isotropic distribution of matter in the universe on large scales. However, recent observations, such as the temperature anisotropy of the cosmic microwave background (CMB) radiation, the motion of galaxies in the universe, the polarization of quasars and the acceleration of the cosmic expansion, indicate preferred directions in the sky. If these directions have a cosmological origin, the cosmological principle would be violated, and modern cosmology should be reconsidered. In this paper, by considering the preferred axis in the CMB parity violation, we find that it coincides with the preferred axes in CMB quadrupole and CMB octopole, and they all align with the direction of the CMB kinematic dipole. In addition, the preferred directions in the velocity flows, quasar alignment, anisotropy of the cosmic acceleration, the handedness of spiral galaxies, and the angular distribution of the fine-structure constant are also claimed to be aligned with the CMB kinematic dipole. Since CMB dipole was confirmed to be caused by the motion of our local group of galaxies relative to the reference frame of the CMB, the coincidence of all these preferred directions hints that these anomalies have a common origin, which is not cosmological or due to a gravitational effect. The systematical or contaminative errors in observation or in data analysis, which can be directly related to the motion of our local group of galaxies, can play an important role in explaining the anomalies.

\end{abstract}


\maketitle

\section{Introduction}

Various cosmological observations support the standard cosmological model: inflation+$\Lambda$CDM model \cite{cmb-review}, including the temperature and polarization anisotropies of the cosmic microwave background (CMB) radiation, the distribution of large-scale structure, Type Ia supernovae, the baryon acoustic oscillation and the cosmic weak lensing. This successful model is based on the following assumptions: (1) The universe is homogeneous and isotropic in large scales (2) Gravity is correctly described by general relativity in all macroscopic scales. (3) Random and nearly Gaussian distributed cosmic anisotropies originated from the quantum fluctuations in the early inflationary stage.

However, the increasing release of precise data, in particular the Wilkinson Microwave Anisotropy Probe (WMAP) and the Planck satellites measurements of the temperature anisotropies of the CMB \cite{wmap,planck}, led to the claim of a number of anomalies in the CMB largest scales: the low quadrupole problem \cite{cobe}, the lack of both variance and correlation on the largest angular scales \cite{lack}, cold spot problem \cite{cold-spot}, power asymmetry \cite{power}, hemisphere asymmetry \cite{hemisphere}, large-scale quadrant asymmetry \cite{quadrant}, alignment of low multipoles \cite{alignment0,alignment,alignment2}, parity asymmetry \cite{parity1}, mirror-parity asymmetry, and so on (see \cite{1001.4613} as the review of WMAP results, \cite{1303.5083,1506.07135} for a review on Planck results, and \cite{review} for a recent review).

Based on these observations, the Planck collaboration stated that ``\emph{The Universe is still weird and interesting}". The origin of these anomalies are still not well understood, and it could hint to the physics of the earliest stage of the universe, preceding the big bang. Other more prosaic explanations are also possible, including foreground microwave emissions from objects that are not yet known and have not been predicted. If the anomalies have a cosmological origin, violating the cosmological principle, the standard model of cosmology should be revised. On the contrary, if they have non-cosmological origins (possible foreground residuals or unsolved systematics), the non-cosmological artifact should be well studied and removed from the data to avoid misleading physical explanations of the universe. So, for any of the above cases, these large-scale anomalies deserve to be well studied.

Among all the anomalies, several ones are direction dependent. For instance, the alignment of the CMB low multipoles, for example, from $\ell=2$ to $\ell=5$ seem to be pointing to a common direction. If the explanation for this anomaly is cosmological, it indicates that there is a preferred direction in our universe, which is a significant violation of the cosmological principle.

Another direction dependent problem is related to the CMB parity asymmetry. This problem has been investigated in the literature \cite{parity1,kim2011} and shows a significant dominance on the CMB power spectrum stored in the odd multipoles over the even ones. This odd parity preference was  also confirmed in the recent Planck data \cite{planck2013,planck2015}. By defining various directional statistics, in previous works \cite{zhao2012,zhao2014,zhao2015}, we investigated the directional properties of the CMB parity asymmetry, and found that CMB parity violation favors a preferred direction, which is independent of the choice of the statistics. In this paper, we shall review the directional properties of the CMB parity asymmetry, and search for the preferred directions stored in CMB low multipoles. In particular, we compare this preferred direction with the announced preferred directions in CMB quadrupole and octopole, and find that these directions have strong correlations. The alignment between them is also confirmed at more than $3\sigma$ confidence level. This result indicates that the CMB parity asymmetry should have a common explanation with other anomalies including the alignment problem of CMB low multipoles, the low quadrupole problem, and the lack of large-scale correlation. Most importantly, we find that all these preferred directions are coincident with the direction of CMB kinematic dipole. In addition, the preferred directions were also reported in a number of other cosmological observations: the velocity flows \cite{velocity}, quasar alignment \cite{quasar}, anisotropy of the cosmic acceleration \cite{acceleration,acc2}, the handedness of spiral galaxies \cite{spiral}, and angular distribution of the fine-structure constant \cite{fine}. Even though there are many debates \cite{debate1,debate2,debate3,debate4,debate5}, it was also reported that all these preferred directions seem to coincide with the CMB kinematic dipole.

It is well known that the CMB kinematic dipole is caused by the motion of our local group of galaxies (including the Milky Galaxies) relative to the reference frame of CMB in the direction of the Galactic coordinate ($\theta=42^{\circ}$, $\phi=264^{\circ}$), which is a pure kinematic effect. So, if the preferred direction of any claimed CMB anomaly coincides with the CMB dipole direction, it should have a non-cosmological origin. The explanation should consider the possible CMB dipole-related foreground residuals or systematical errors that should be seriously handled in the future measurements, and which could hint to some unsolved contaminations in the large-scale observations, including CMB, galaxies distribution, Type Ia supernovae, quasar distribution, and so on.

The rest of this article will be structured as follows. In Sec. 2, we briefly summarize the anomaly on CMB parity asymmetry. In Sec. 3, we focus on the directional properties and the preferred directions of CMB parity asymmetry. In Sec. 4, we compare this preferred direction with the preferred directions stored in the CMB dipole, quadrupole and octopole. In Sec. 5, we briefly introduce the other large-scale anomalies and their direction dependence. In Sec. 6, we list some possible explanations for the cosmological anomalies mentioned above. Sec. 7 is contributed as a summary and conclusion of this paper.

\section{CMB parity asymmetry}

The CMB temperature fluctuation on a two-dimensional sphere is a scalar field. According to the coordinate transformation, it can be decomposed as the standard spherical harmonics as follows,
\begin{equation}
\Delta T(\hat{n})=\sum_{\ell=0}^{\infty} \sum_{m=-\ell}^{\ell} a_{\ell m} Y_{\ell m} (\hat{n}),
\end{equation}
where $Y_{\ell m}(\hat{n})$ are the spherical harmonics, and $a_{\ell m}$ are the corresponding coefficients. In the standard inflationary scenario, both primordial scalar and tensor perturbations are random Gaussian fields. In the linear order approximation, the two-dimensional temperature fluctuations also satisfy the random Gaussian distribution, i.e., the amplitudes $|a_{\ell m}|$ are distributed according to Rayleigh's probability distribution function, and the phase of $a_{\ell m}$ with $m\neq 0$ is supported to be evenly distributed in the range $[0,2\pi]$. For the random Gaussian field, the statistical properties are completely described by the second-order power spectrum, namely
\begin{equation}
\langle a_{\ell m}a^*_{\ell' m'}\rangle=C_{\ell}\delta_{\ell\ell'}\delta_{mm'},
\end{equation}
where $\langle ... \rangle$ denotes the average over the statistical ensemble of realizations, and the spectrum $C_{\ell}$ is independent of the magnetic quantum number $m$ for the statistical isotropic field. In the real measurements, it is impossible to directly observe the power spectrum $C_{\ell}$ itself. One has to construct the estimators. For the full-sky map, if  the noise is negligible, the best unbiased estimator for $C_{\ell}$ is \cite{grishchuk1997}
\begin{equation}\label{hat-cl}
\hat{C}_{\ell}=\frac{1}{2\ell+1} \sum_{m=-\ell}^{\ell} a_{\ell m} a_{\ell m}^*,
\end{equation}
and the statistical uncertainty is $\Delta \hat{C}_{\ell}=\sqrt{\frac{2}{2\ell+1}}C_{\ell}$, which is the so-called cosmic variance.

\begin{figure}[t]
\begin{center}
\includegraphics[width=14cm]{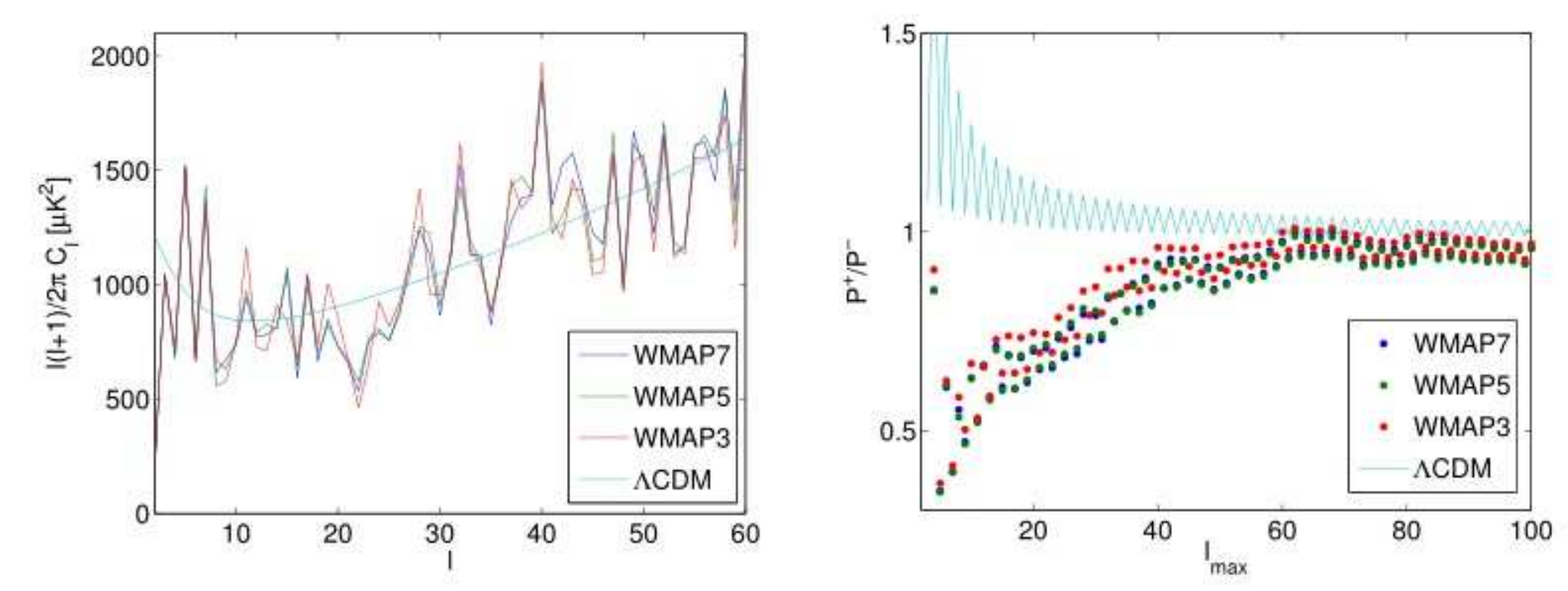}
\end{center}
\caption{Left panel: The predicted CMB low multipole power spectrum in the $\Lambda$CDM model, and 3-year, 5-year and 7-year WMAP observed data; Right panel: Theoretical values of $P^+/P^-$ compares with the WMAP observed results \cite{1202.0728}.}\label{fig1}
\end{figure}

According to the symmetry under the transformation of coordinate inversion $\hat{n}\rightarrow -\hat{n}$, CMB anisotropy field can be decomposed as the symmetric component $\Delta T^{+}$ and the antisymmetric component $\Delta T^{-}$ as follows,
\begin{eqnarray}
\Delta T^{+}(\hat{n})=\frac{\Delta T(\hat{n})+\Delta T(-\hat{n})}{2}, ~~
\Delta T^{-}(\hat{n})=\frac{\Delta T(\hat{n})-\Delta T(-\hat{n})}{2}.
\end{eqnarray}
The patterns $\Delta T^{+}(\hat{n})$ and $\Delta T^{-}(\hat{n})$ have even and odd-parity respectively, which can also be written as
\begin{eqnarray}
\Delta T^{+}(\hat{n})=\sum_{\ell m} a_{\ell m} Y_{\ell m}(\hat{n})\Gamma_{\ell}^{+}, ~~
\Delta T^{-}(\hat{n})=\sum_{\ell m} a_{\ell m} Y_{\ell m}(\hat{n})\Gamma_{\ell}^{-},
\end{eqnarray}
where $\Gamma_{\ell}^{+}=\cos^2\left(\frac{\ell\pi}{2}\right)$ and $\Gamma_{\ell}^{-}=\sin^2\left(\frac{\ell\pi}{2}\right)$.
Therefore, significant power asymmetry between even and odd multipoles may be interpreted as a preference for a particular parity of the anisotropy pattern. Fig. \ref{fig1} (left panel) presents the observed low multipole data by WMAP satellite. Comparing with the theoretical predictions, we find that in the lowest multipole range, the odd multipoles are systematically larger than  expected, while the even multipoles are systematically smaller than the theoretical ones. This regular pattern significantly violates the random and Gaussian assumption of the CMB field, and strongly indicates the odd-multipole preference, i.e. the antisymmetric preference of the CMB anisotropy.

In order to quantify this asymmetry, we introduce the following statistic:
\begin{eqnarray}
P^{+}=\sum_{\ell=2}^{\ell_{\max}}\frac{\ell(\ell+1)}{2\pi}C_{\ell}\Gamma_{\ell}^{+},
~~P^{-}=\sum_{\ell=2}^{\ell_{\max}}\frac{\ell(\ell+1)}{2\pi}C_{\ell}\Gamma_{\ell}^{-}.
\end{eqnarray}
$P^{+}$ and $P^{-}$ are the sum of the power spectrum for even and odd multipoles, respectively. Therefore, the ratio $P^{+}/P^{-}$ is associated with the degree of the parity asymmetry, where the lower value of $P^{+}/P^{-}$ indicates the odd-parity preference, and vice-versa. Fig. \ref{fig1} (right panel) shows the ratio derived from the WMAP observed data. Comparing with the theoretical values, we find that if the maximum multipole $\ell_{\max}$ is small, i.e., in the low multipole range, the observed results significantly deviate from the model predictions, showing an obvious odd-multipole preference. This is the so-called CMB parity asymmetry anomaly.

\begin{figure}[t]
\begin{center}
\includegraphics[width=14cm]{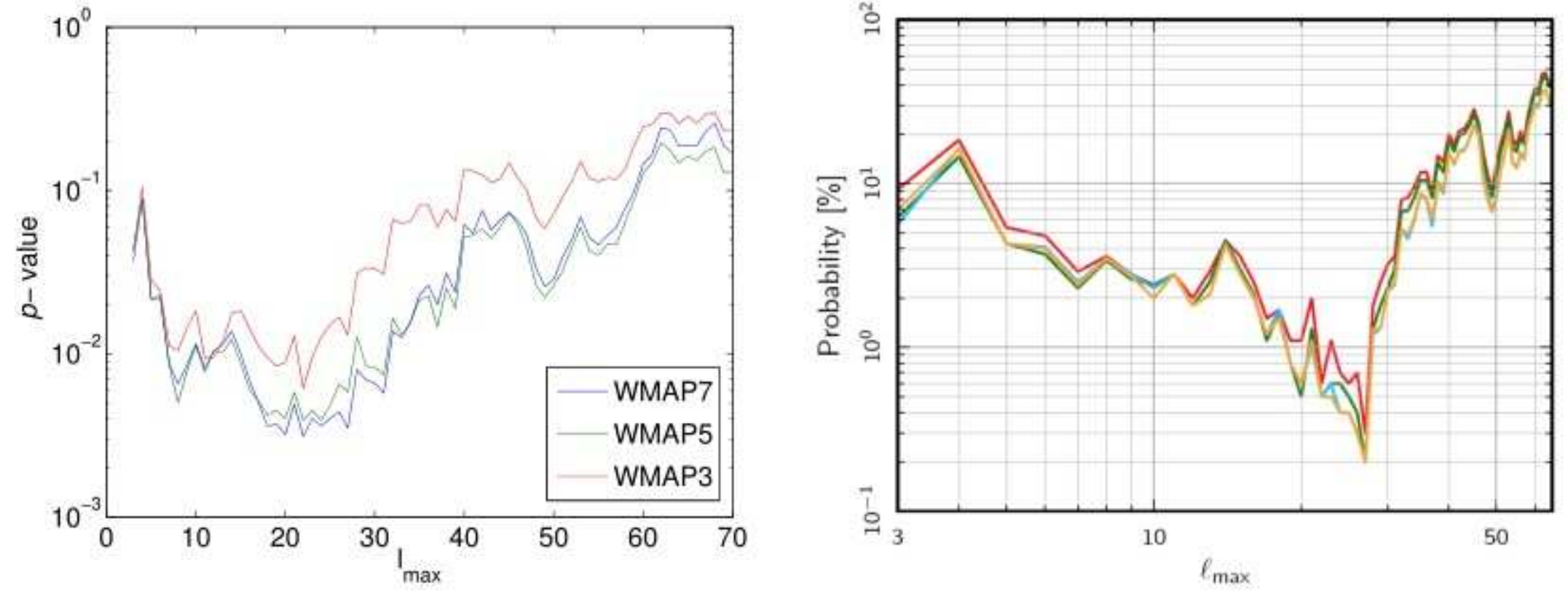}
\end{center}
\caption{Left panel: Probability of getting $P^+/P^-$ as low as WMAP data for multipole range $2\le\ell\le\ell_{\max}$ \cite{1202.0728}. Right panel: Probability of getting $P^+/P^-$ as low as Planck Commander (red), NILC (orange), SEVEM (green), SMICA (blue) data for multipole range $2\le\ell\le\ell_{\max}$ \cite{1506.07135}.}\label{fig2}
\end{figure}

In order to quantify the statistical significance of this anomaly, as in general, we can simulate a large number of random Gaussian samples based on the best-fit $\Lambda$CDM model. For each sample, we calculate the ratio $P^{+}/P^{-}$, and count the probability (i.e. the $p$-value) to get the ratio $P^{+}/P^{-}$, which is smaller than the observed value for each $\ell_{\max}$. In Fig. \ref{fig2} (left panel), we plot the $p$-values for various $\ell_{\max}$, from which we find that the $p$-value curves minimize at $\ell_{\max}\sim (20,30)$, and the minimal $p$-value is less than $1\%$ in the multipole range. Recently, Planck collaboration repeated this calculation using the new released data. The corresponding curves for the $p$-values as function of $\ell_{\max}$ are shown in Fig. \ref{fig2} (right panel), from which we find the results derived from Planck data are quite close to those from the WMAP data, and all of them indicate a significant odd-parity preference in the CMB low multipoles, and the minimal $p$-value is much smaller than $1\%$ for all the Planck released data. So, we conclude that the CMB low-multipole data indicate a significant violation of parity symmetry, which conflicts with the prediction of the random Gaussian assumption of the standard cosmological model.


\begin{figure}[t]
\begin{center}
\includegraphics[width=10cm]{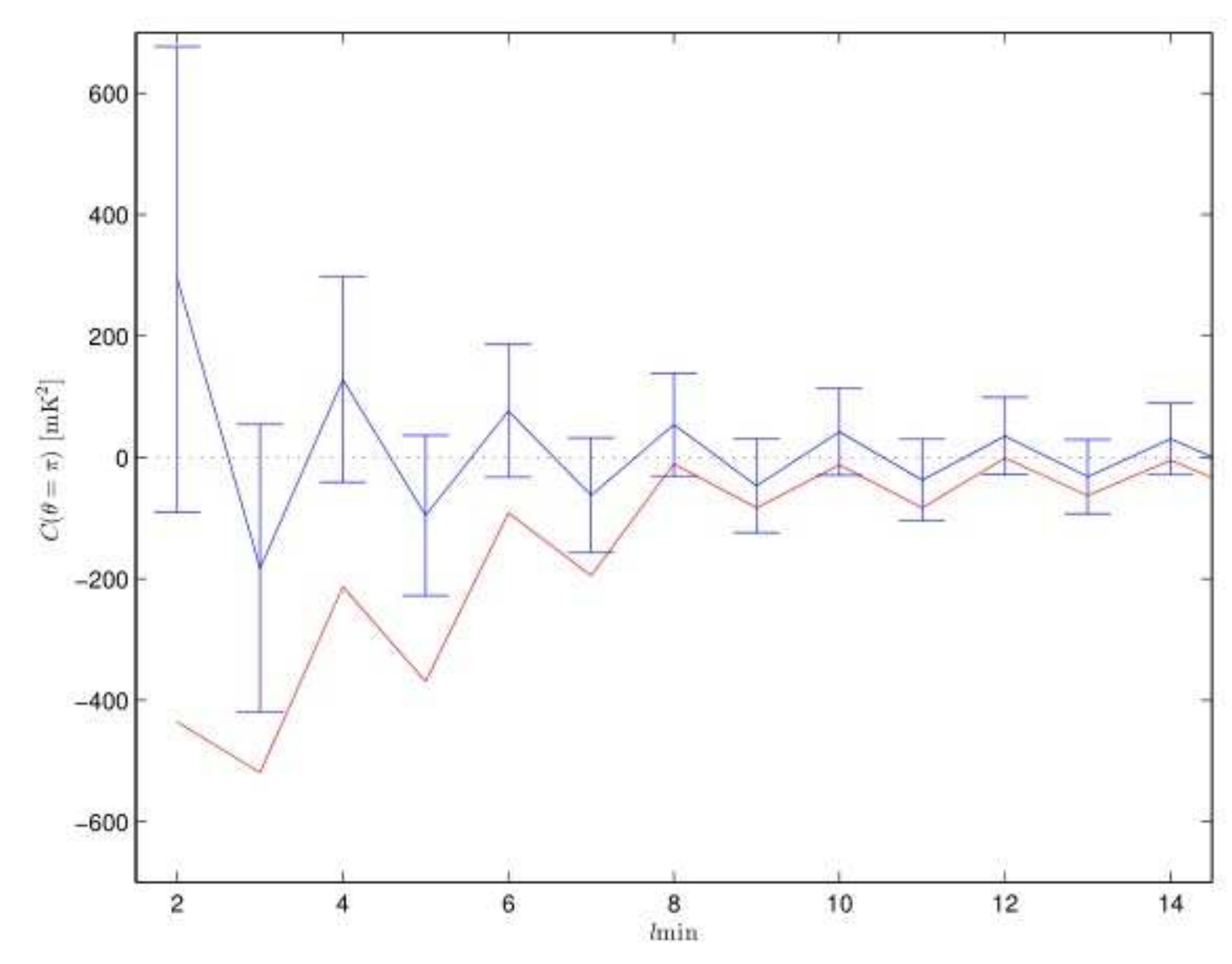}
\end{center}
\caption{Theoretical (blue curve) and observed (red curve) values of $C(\Theta=\pi)$ as a function of $\ell_{\min}$, where the seven-year WMAP power spectrum have been used as the observed data. The error bar indicates the $1\sigma$ confident level caused by cosmic variance \cite{zhao2012}.}\label{fig4}
\end{figure}

The parity asymmetry problem can also be understood by studying the temperature correlation between any direction and its opposite one. For the two-dimensional spherical CMB map, the two-point correlation function is given be
\begin{equation}\label{C_theta}
C(\Theta)=\langle \Delta T(\hat{n})\Delta T(\hat{n'})\rangle=\sum_{\ell=\ell_{\min}}^{\infty}\frac{2\ell+1}{4\pi} C_{\ell} P_{\ell}(\cos\Theta),
\end{equation}
where $P_{\ell}$ are the Legendre polynomials, and $\cos\Theta=\hat{n}\cdot\hat{n'}$. From this formula, we can easily show the correlations of the largest angular distance,
\begin{equation}\label{C_phi}
C(\Theta=\pi)=\sum_{\ell=\ell_{\min}}^{\infty}\frac{2\ell+1}{4\pi} C_{\ell} (\Gamma_{\ell}^{+}-\Gamma_{\ell}^{-}).
\end{equation}
For the even-parity preference case, we have a positive correlation, and for the odd-parity preference case, we have an anti-correlation. In Fig. \ref{fig4}, we plot the theoretical prediction of the correlation function $C(\Theta=\pi)$ for different $\ell_{\min}$, and also compare them with the results derived from the real WMAP data. Clearly, we find that the model predictions favor the even-parity preference, while the real data seems favor the odd-parity preference. So, this analysis also shows an odd-parity preference of CMB low-multipole data, which is consistent with the conclusion above.

\section{Preferred axis of CMB parity violation}

\subsection{Preferred axis in the full-sky maps}

In this paper, we shall mainly investigate the directional properties of the CMB parity asymmetry. In order to realize it, we need a directional dependent statistic. However, as shown in the previous section, all the statistics defined to describe the parity problem are based on the power spectra $C_{\ell}$ and their estimators $\hat{C}_{\ell}$, which are all rotationally invariant. Statistical invariance means that for any rotation of the reference system of coordinate, the power spectrum and the correlation function are invariant. So, any statistic defined by them are all coordinate independent. In order to break this kind of invariance, similar to other works \cite{alignment0,alignment,alignment2}, we should replace the estimator $\hat{C}_{\ell}$ by rotationally variant estimators. First, let us work on the full-sky CMB maps. The simplest one can be defined as following,
\begin{equation}\label{D_l}
\hat{D}_{\ell}=\frac{1}{2\ell}\sum_{m=-\ell}^{\ell} a_{\ell m} a_{\ell m}^*(1-\delta_{m0}),
\end{equation}
where $\delta_{mm'}$ is the Kroneker symbol. From the definition, we know that $\hat{D}_{\ell}$ is also an unbiased estimator for the power spectrum $C_{\ell}$ for any given multipole, i.e. $\langle \hat{D}_{\ell}\rangle=C_{\ell}$. Now, we can study the estimator $\hat{D}_{\ell}$ in any coordinate system. Imagining that the Galactic coordinate system is rotated by the Euler angle $(\psi,\theta,\phi)$, the coefficients $a_{\ell m}(\psi,\theta,\phi)$ in this new coordinate system can be calculated by
\begin{equation}
a_{\ell m} =\sum_{m=-\ell}^{\ell} a_{\ell m'} D_{m m'}^{\ell}(\psi,\theta,\phi),
\end{equation}
where $a_{\ell m}\equiv a_{\ell m}(0,0,0)$ are the coefficients defined in the Galactic coordinate system, and $D_{m m'}^{\ell}(\phi,\theta,\phi)$ is the Wigner rotation matrix \cite{edmond}. Similar to Eq. (\ref{D_l}), we can define the estimator $\hat{D}_{\ell}(\psi,\theta,\phi)$. It is easy to find that $\hat{D}_{\ell}(\psi,\theta,\phi)$ is independent of the angle $\psi$. So in this paper, we only consider two Euler angle $\hat{\rm{\bf q}}\equiv(\theta,\phi)$ and set $\psi=0$. If we consider $\hat{\rm{\bf q}}$ as a vector, which labels the $z$-axis direction in the rotated coordinate system, then $(\theta,\phi)$ is the polar coordinate of this direction in the Galactic system, which relates to the Galactic coordinate $(l,b)$ by $b=90^{\circ}-\theta$ and $l=\phi$. For any given coordinate labeled by $\hat{\rm{\bf q}}$, the components $a_{\ell 0}$ are naturally symmetric around the $z$-axis, i.e. $\hat{\rm{\bf q}}$. \emph{So from the definition of $\hat{D}_{\ell}$, in which the $m=0$ components are excluded, we know, in this coordinate system, that the $z$-axis (i.e. $\hat{\rm{\bf q}}$) is the preferred axis. If we rotate the coordinate system, the preferred axis also rotates.}

Now, we can define the rotationally variable parity parameter $G_1(\ell;\hat{\rm{\bf q}})$ as follows,
\begin{equation}\label{G_1}
G_1(\ell;\hat{\rm{\bf q}})=\frac{\sum_{\ell'=2}^{\ell}{\ell'(\ell'+1)}\hat{D}_{\ell'}(\hat{\rm{\bf q}})\Gamma_{\ell'}^{+}}{\sum_{\ell'=2}^{\ell}{\ell'(\ell'+1)}\hat{D}_{\ell'}(\hat{\rm{\bf q}})\Gamma_{\ell'}^{-}}.
\end{equation}
This statistic stands for the amplitude of the original parity parameter $P^+/P^-$, which is associated with the degree of the parity asymmetry, where a value of $G_1<1$ indicates an odd-parity preference, and $G_1>1$ indicates an even-parity preference.  At the same time, due to the rotational variance of $G_1(\ell;\hat{\rm{\bf q}})$, we can study the possible preferred direction, which may reveal hints on the origin of the observed parity asymmetry in the CMB field. For any given $\ell$, the sky map $G_1(\ell;\hat{\rm{\bf q}})$ can be constructed by considering all directions $\hat{\rm{\bf q}}$. In practice, we pixelize the full sky in HEALPix format with the resolution parameter $N_{\rm side}=64$ and set the direction $\hat{\rm{\bf q}}$ to be those of the pixels.

\begin{figure*}[t]
\begin{center}
\includegraphics[width=15cm]{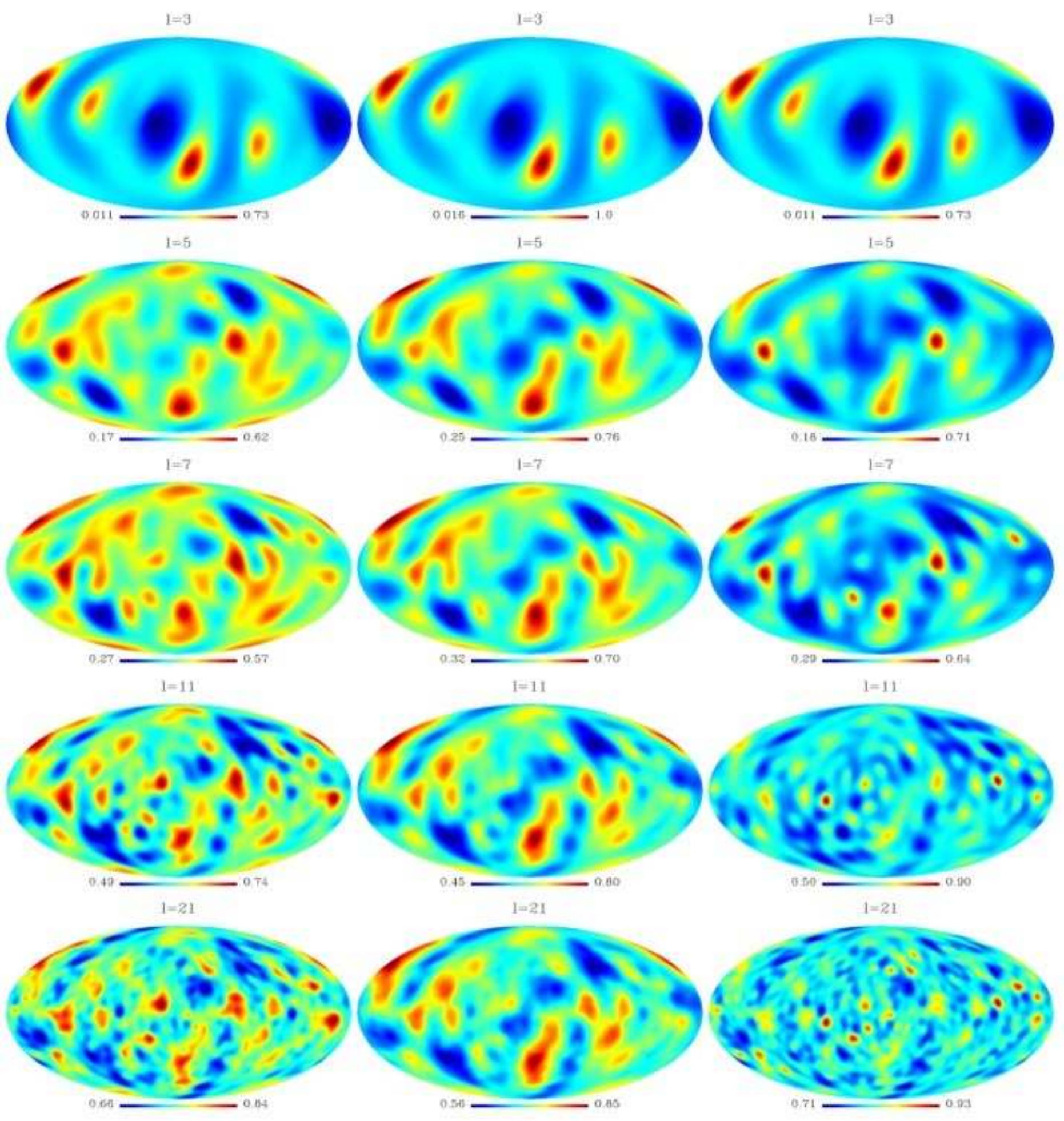}
\end{center}
\caption{Three directional statistics $G_{1}(\ell;\hat{\bf q})$ (left),
$G_2(\ell;\hat{\bf q})$ (middle), and $G_3(\ell;\hat{\bf q})$ (right) as
functions of $\hat{\bf q}\equiv (\theta,\phi)$. Note that,
these results are based on the Planck 2013 SMICA data \cite{zhao2014}.}\label{fig5}
\end{figure*}

{In previous works, the authors in \cite{kim2011}, using the power spectrum $C_{\ell}$, found that the CMB parity asymmetry is quite significant at the low multipoles, and this tendency extends to the multipole range $\ell<22$. Since the definition of the estimator $\hat{D}_{\ell}$ is similar to that of $C_{\ell}$, one expects the parity asymmetry of the statistic $G_1$ to also extend to this multipole range.} Using the Planck 2013 SMICA map, we compute the directional parity parameter $G_1(\ell;\hat{\rm{\bf q}})$ for any direction $\hat{\rm{\bf q}}$. As we have mentioned, $\hat{\rm{\bf q}}$ labels the $z$-axis direction in the rotated  coordinate system, and $(\theta,\phi)$ is just the the polar coordinate of this direction in the Galactic coordinate. We plot the parameter $G_1(\ell;\hat{\rm{\bf q}})$ as a function of $\hat{\rm{\bf q}}$ for $3\le\ell\le22$ in Fig. \ref{fig5} (left panels). From this figure, we find that $G_1(\ell;\hat{\rm{\bf q}})<1$ holds for any direction $\hat{\rm{\bf q}}$ and maximum multipole $\ell$, which is consistent with the discovery of the odd-parity preference in the previous discussions.  Smaller $G_1$ value leads to larger parity violation. In addition, we find that for any given maximum multipole $\ell$, except for the case with $\ell=3$, all the $G_1$ maps have quite similar morphologies. In Table \ref{tab2}, we list the preferred direction $\hat{\rm{\bf q}}$, where the parity parameter $G(\ell;\hat{\rm{\bf q}})$ for each maximum $\ell$ is minimized, and find that in all these cases, the preferred directions are nearly the same. Note that, throughout this paper, we do not differentiate the direction $\hat{\rm{\bf q}}$ and the opposite one $-\hat{\rm{\bf q}}$.

\begin{table*}
\caption{The preferred direction $\hat{\bf q}=(\theta,\phi)$,
where the parity parameter $G_i(\ell;\hat{\bf q})$ based on Planck 2013
SMICA data is minimized, compared with the other CMB preferred
axes. In each box, the upper one is the result for the statistic with $i=1$,
the middle one is that for $i=2$, and the lower one is that for $i=3$. In
this table, $\alpha$ is the angle between $\hat{\bf q}$ and the
CMB kinematic dipole, $\langle|\cos\theta_{ij}|\rangle$ is the
quantity defined in Eq. (\ref{ctheta}), and the $\Delta_c/\sigma_c$ value denotes
the number of $\sigma_c$ the observed
$\langle|\cos\theta_{ij}|\rangle$ deviates from the simulations \cite{zhao2014}.}
\begin{center}
\label{tab2}
\begin{tabular}{ |c||c |c |c |c |c| }
    \hline
     & ~~~~~~$\theta[^{\circ}]$~~~~~~ & ~~~~~~$\phi[^{\circ}]$~~~~~~ & ~~~~~~$|\cos\alpha|$~~~~~~ & ~~~~$\langle|\cos\theta_{ij}|\rangle$~~~~ & ~~~~~~~$\Delta_c/\sigma_c$~~~~~~~  \\
   \hline
   \hline
   \multirow{3}{*}{$\ell_{\max}=3$} &    90.00  &  23.20    & 0.3265 &    0.6066        & 0.90 \\
   &    90.00  &  23.20    & 0.3265 &    0.6066        & 0.90 \\
   &    90.00  &  23.20    & 0.3265 &    0.6066        & 0.90 \\
   \hline
   \multirow{3}{*}{$\ell_{\max}=5$} &   45.80  &   281.07   & 0.9767 &     0.9015    & 3.40 \\
   &   45.80  &   281.07   & 0.9767 &     0.9015    & 3.40 \\
   &   45.80  &   281.07   & 0.9767 &     0.9015    & 3.40 \\
   \hline
   \multirow{3}{*}{$\ell_{\max}=7$} &   48.19  &   277.73   & 0.9799 &     0.8979    & 3.37 \\
   &   47.39  &   279.29   & 0.9782 &     0.8987    & 3.38 \\
   &   52.83  &   267.89   & 0.9710 &     0.8915    & 3.32 \\
   \hline
   \multirow{3}{*}{$\ell_{\max}=11$} &   52.08  &   284.06   & 0.9525 &     0.8744    & 3.17 \\
   &   49.77  &   280.54   & 0.9697 &     0.8886    & 3.29 \\
   &   53.58  &   226.41   & 0.8679 &     0.8793    & 3.21 \\
   \hline
   \multirow{3}{*}{$\ell_{\max}=21$} &   52.08  &   285.47   & 0.9479 &     0.8721    & 3.15 \\
   &   50.55  &   284.06   & 0.9575 &     0.8804    & 3.22 \\
   &   21.32  &   131.90   & 0.5292 &     0.8295    & 2.79 \\
   \hline
\end{tabular}
\end{center}
\end{table*}

\begin{table}
\caption{The definitions of six directional statistics considered in the text.}
\begin{center}
\label{tab1}
\begin{tabular}{ |c|c|  }
   \hline
   Number of statistic  & Definition \\
   \hline
   $1^{\rm st}$  &    $G_1(\ell;\hat{\bf q})$ with $\hat{D}_{\ell}(\hat{\bf q})$       \\
      \hline
   $2^{\rm nd}$  &    $G_2(\ell;\hat{\bf q})$ with $\hat{D}_{\ell}(\hat{\bf q})$       \\
      \hline
   $3^{\rm rd}$  &    $G_3(\ell;\hat{\bf q})$ with $\hat{D}_{\ell}(\hat{\bf q})$       \\
      \hline
   $4^{\rm th}$  &    $G_1(\ell;\hat{\bf q})$ with $\hat{D}'_{\ell}(\hat{\bf q})$       \\
      \hline
   $5^{\rm th}$  &    $G_2(\ell;\hat{\bf q})$ with $\hat{D}'_{\ell}(\hat{\bf q})$       \\
      \hline
   $6^{\rm th}$  &    $G_3(\ell;\hat{\bf q})$ with $\hat{D}'_{\ell}(\hat{\bf q})$       \\
   \hline
\end{tabular}
\end{center}
\end{table}

\subsection{Independence of the statistics}

In the previous discussion, based on the analysis of the parity parameter $G_1(\ell;\hat{\rm{\bf q}})$, we found the preferred direction of the CMB parity asymmetry, which is independent of the maximum multipole $\ell$. However, an important problem arises: Whether or not the conclusion derived above depends on the definition of the statistic or estimator? In order to cross-check the result, we consider another rotationally variant estimator, which is proposed by de Oliveira-Costa et al. \cite{de2004}
\begin{equation}
\hat{D}'_{\ell}\equiv \frac{1}{2\ell+1}\sum_{m=-\ell}^{\ell} m^2 |a_{\ell m}|^2.
\end{equation}
If our universe is statistically isotropic, the ensemble average of this estimator is related to the power spectrum as follows,
\begin{equation}
\langle \hat{D}'_{\ell}\rangle =\frac{\ell(\ell+1)}{3}C_{\ell}.
\end{equation}
\emph{As we discussed, this estimator has also chosen a preferred direction, i.e. the $z$-axis direction.} In addition, this estimator favors high $m$s and so it works well in searches for planarity. In a quantum-mechanical system, this quantity also corresponds to the angular momentum along the $z$-axis direction. So the statistic defined by this estimator can also be used to search for the preferred axis in the CMB field.

In order to avoid the dependence of the statistic, in addition to the estimator $G_1(\ell;\hat{\rm{\bf q}})$ defined in Eq. (\ref{G_1}), we also consider the following different statistics, which are constructed by using two kinds of estimators $\hat{D}_{\ell}$ and $\hat{D}'_{\ell}$. From the definition of the two-point function in Eqs. (\ref{C_theta}) and (\ref{C_phi}), we can easily define the estimator for the correlation function at the largest angular distance $\Theta=\pi$ as,
\begin{equation}
\hat{C}(\Theta=\pi;\hat{\rm{\bf q}})=\sum_{\ell}\frac{(2\ell+1)}{4\pi} \hat{D}_{\ell}(\hat{\rm{\bf q}})(\Gamma^{+}_{\ell}-\Gamma^{-}_{\ell}).
\end{equation}
So, the natural way to estimate the relative contribution of the even and odd multipoles to the correlation function is to define the statistic
\begin{equation}\label{G_2}
G_2(\ell;\hat{\rm{\bf q}})=\frac{\sum_{\ell'=2}^{\ell}{(2\ell'+1)}\hat{D}_{\ell'}(\hat{\rm{\bf q}})\Gamma_{\ell'}^{+}}{\sum_{\ell'=2}^{\ell}{(2\ell'+1)}\hat{D}_{\ell'}(\hat{\rm{\bf q}})\Gamma_{\ell'}^{-}},
\end{equation}
which follows that $\hat{C}(\Theta=\pi)\propto (G_2(\ell;\hat{\rm{\bf q}})-1)$. $G_2>1$ corresponds to the positive correlation of the opposite direction, and $G_2<1$ indicates the anticorrelation of them. Note that the statistic $G_2$ is different from $G_1$, due to the different factors before $\hat{D}_{\ell}$ in their definitions. Therefore, the relative weights of low multipoles are much higher in $G_2$ than those in $G_1$.

For further investigation, in this paper, we also consider a third statistic to quantify the parity asymmetry, which was first introduced in Ref.\cite{Aluri2012}
\begin{equation}\label{G_3}
G_3(\ell;\hat{\rm{\bf q}})=\frac{2}{\ell-1}\sum_{\ell'=3}^{\ell} \frac{(\ell'-1)\ell'\hat{D}_{\ell'-1}(\hat{\rm{\bf q}})}{\ell'(\ell'+1)\hat{D}_{\ell'}(\hat{\rm{\bf q}})},
\end{equation}
where the maximum $\ell$ is any odd multipole $\ell\ge 3$ and the summation is over all odd multipoles up to $\ell$. This statistic is the measure of the mean deviation of the ratio of power in the even multipole and its succeeding odd-multipole from one.

We apply these two statistics for all the odd multipoles $3\le \ell
\le 21$ to the released Planck 2013 SMICA, NILC and SEVEM data. The results for SMICA
data are presented in Fig. \ref{fig5} (middle and right panels). For all the odd maximum multipoles $\ell$
and directions $\hat{\bf q}$, we have $G_i<1$ for $i=2,3$. These
are also correct for both Planck NILC and SEVEM data. So, we
find that the real CMB data have the odd-parity preference, which
is independent of the choice of the parity statistics.

The other three directional statistics are defined by similar manner, but the estimator $\hat{D}_{\ell}$ is replaced by $\hat{D}'_{\ell}$, i.e.
\begin{equation}
G_4(\ell;\hat{\rm{\bf q}})=G_1(\ell;\hat{\rm{\bf q}})|_{\hat{D}_{\ell}\rightarrow\hat{D}'_{\ell}},
\end{equation}
\begin{equation}
G_5(\ell;\hat{\rm{\bf q}})=G_2(\ell;\hat{\rm{\bf q}})|_{\hat{D}_{\ell}\rightarrow\hat{D}'_{\ell}},
\end{equation}
\begin{equation}
G_6(\ell;\hat{\rm{\bf q}})=G_3(\ell;\hat{\rm{\bf q}})|_{\hat{D}_{\ell}\rightarrow\hat{D}'_{\ell}},
\end{equation}
which are summarized in Table \ref{tab1}.

\begin{figure*}[t]
\begin{center}
\includegraphics[width=15cm]{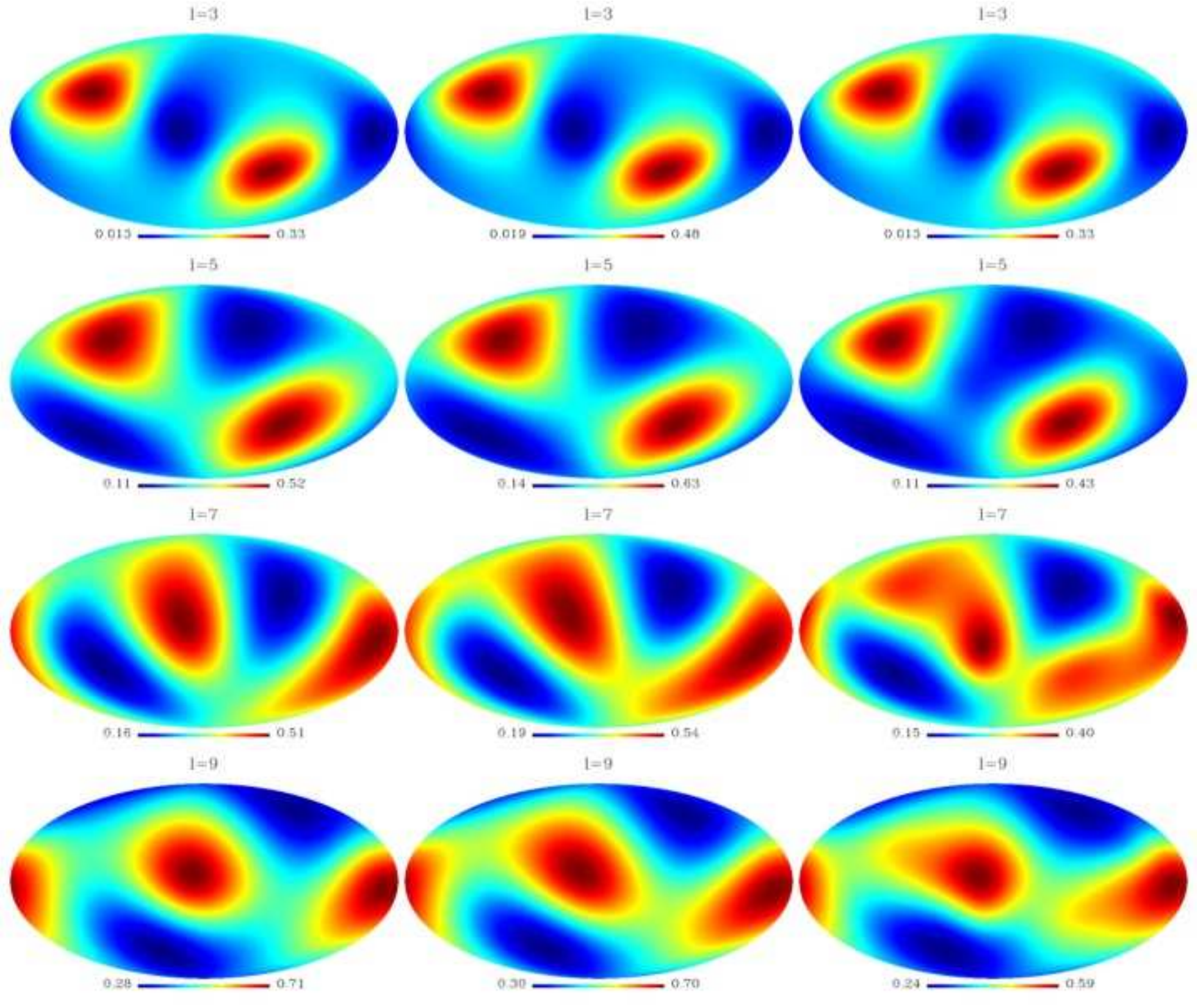}
\end{center}
\caption{Three directional statistics $G_4(\ell;\hat{\bf q})$ (left),
$G_5(\ell;\hat{\bf q})$ (middle), and $G_6(\ell;\hat{\bf q})$ (right) as
functions of $\hat{\bf q}\equiv (\theta,\phi)$. Note that
these results are based on the Planck 2013 SMICA data \cite{zhao2014}.}\label{fig6}
\end{figure*}

{As we have emphasized, the estimator $\hat{D}'_{\ell}$ is quite different from $\hat{D}_{\ell}$ due to the factor $m^2$ in the definition. In the statistics $G_4$, $G_5$, or $G_6$, the contributions of the higher multipoles, $\ell\sim \ell_{\max}$, become completely dominant. For this reason, we only apply these statistics to the multipole range in which the CMB parity asymmetry is most obvious. Although the CMB parity asymmetry can extend to multipole ranges up to $\ell \sim 22$, the main contribution comes from the lowest multipoles, i.e., $\ell<10$, which can be clearly seen in Fig. \ref{fig4}. So, we only consider the parity statistics $G_4$, $G_5$, and $G_6$ for the multipoles $\ell < 10$.} We apply these three statistics to the Planck 2013 SMICA, NILC, and SEVEM data and find similar results. In Fig.\ref{fig6}, we present the results of SMICA data, which show that $G_i(\ell,\hat{\bf q})< 1$ ($i=4,~5,6~$) is held for any direction $\hat{\bf q}$. Therefore, we conclude that the odd-parity preference exists even if the estimator $\hat{D}'_{\ell}$ is considered. In Tables \ref{tab2} and \ref{tab3}, we list the preferred directions $\hat{\bf q}$ for six statistics, which are quite similar to each other. So, we conclude that the preferred direction in the CMB parity asymmetry is independent of the definition of statistics.

\begin{table*}
\caption{The preferred direction $\hat{\bf q}=(\theta,\phi)$,
where the parity parameter $G_i(\ell;\hat{\bf q})$ based on Planck 2013
SMICA data is minimized, compared with the other CMB preferred
axes. In each box, the upper one is the result for the statistic with $i=4$,
the middle one is that for $i=5$, and the lower one is that for $i=6$. In
this table, $\alpha$ is the angle between $\hat{\bf q}$ and the
CMB kinematic dipole, $\langle|\cos\theta_{ij}|\rangle$ is the
quantity defined in Eq. (\ref{ctheta}), and the $\Delta_c/\sigma_c$ value denotes
the number of $\sigma_c$ the observed
$\langle|\cos\theta_{ij}|\rangle$ deviate from the simulations \cite{zhao2014}.}
\begin{center}
\label{tab3}
\begin{tabular}{|c||c |c |c |c |c| }
    \hline
     & ~~~~~~$\theta[^{\circ}]$~~~~~~ & ~~~~~~$\phi[^{\circ}]$~~~~~~ & ~~~~~~$|\cos\alpha|$~~~~~~ & ~~~~$\langle|\cos\theta_{ij}|\rangle$~~~~ & ~~~~~~~$\Delta_c/\sigma_c$~~~~~~~  \\
   \hline
   \hline
   \multirow{3}{*}{$\ell_{\max}=3$} &    88.81  &  23.20    & 0.3109 &    0.5975        & 0.83 \\
   &    88.81  &  23.20    & 0.3109 &    0.5975        & 0.83 \\
   &    88.81  &  23.20    & 0.3109 &    0.5975        & 0.83 \\
   \hline
   \multirow{3}{*}{$\ell_{\max}=5$} &   47.39  &   307.86   & 0.8582 &     0.8458    & 2.93 \\
   &   47.39  &   310.71   & 0.8408 &     0.8390    & 2.87 \\
   &   46.59  &   309.92   & 0.8488 &     0.8442    & 2.92 \\
   \hline
   \multirow{3}{*}{$\ell_{\max}=7$} &   62.72  &   280.55   & 0.9107 &     0.8306    & 2.80 \\
   &   56.49  &   281.25   & 0.9431 &     0.8599    & 3.05 \\
   &   55.77  &   281.95   & 0.9443 &     0.8621    & 3.07 \\
   \hline
   \multirow{3}{*}{$\ell_{\max}=9$} &   32.60  &   236.25   & 0.9451 &     0.9424    & 3.75 \\
   &   36.43  &   248.88   & 0.9815 &     0.9418    & 3.74 \\
   &   34.89  &   246.06   & 0.9737 &     0.9435    & 3.76 \\
   \hline

\end{tabular}
\end{center}
\end{table*}

\subsection{Independence of the CMB masks}

\begin{figure}[t]
\begin{center}
\includegraphics[width=15cm]{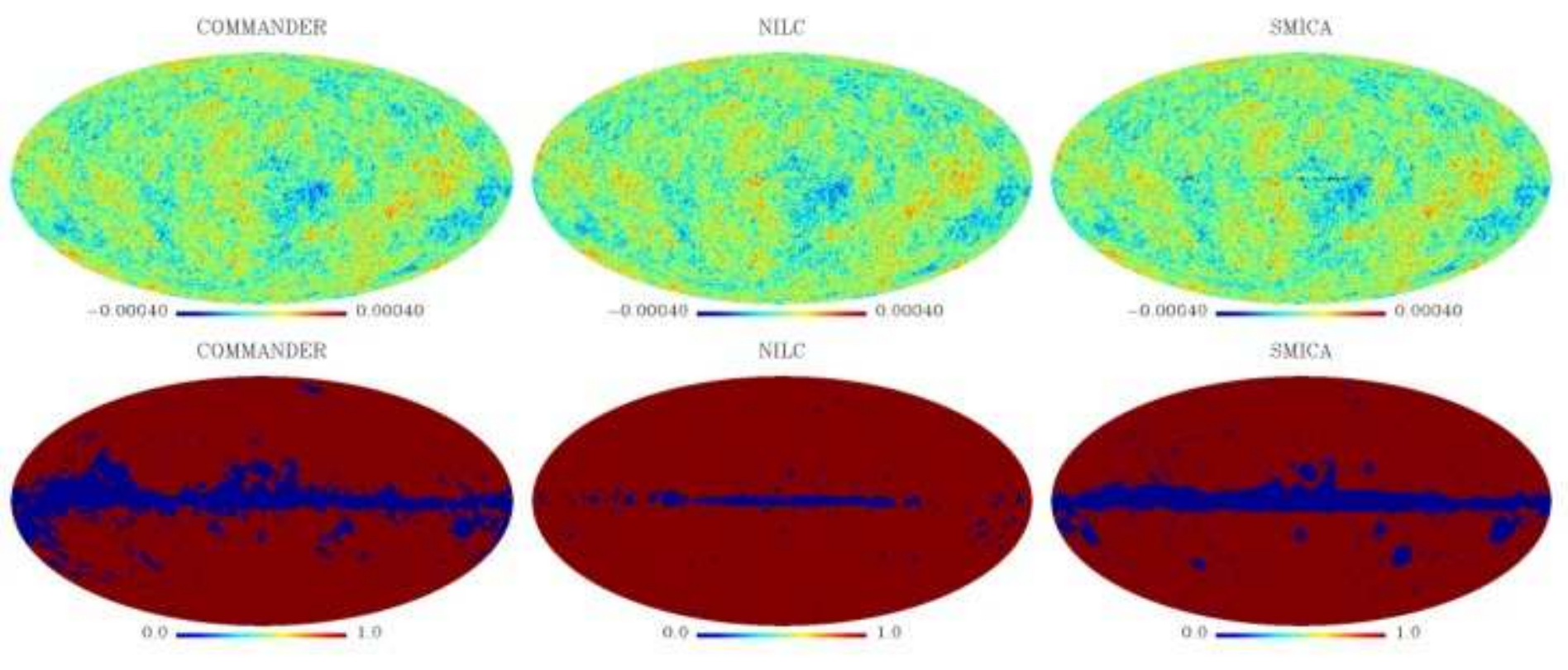}
\end{center}
\caption{The 2015 Planck temperature anisotropy maps, including Commander, NILC, and SMICA. The lower panels are the corresponding masks suggested by Planck collaborations \cite{zhao2015}.}\label{fig7}
\end{figure}

In the CMB observations, various foreground residuals are always unavoidable, especially in the Galactic region. The foreground residuals for the CMB maps released by Planck in 2015 are shown in Fig.\ref{fig7}. Usually one anticipates that the effects of these residuals are small and ignorable in the low multipole range. However, it is still worthy to investigate the cases in which these contaminated data are excluded. The simplest way to exclude the polluted region is to apply the top-hat mask to the data. For each CMB map, the corresponding mask suggested by Planck collaboration is also shown in Fig. \ref{fig7} (lower panels). We find that the masked region in the Commander and SMICA maps are quite similar. While the masked region for the NILC map is quite small, and the information loss in NILC map is expected to be much smaller than in the other two maps. For the masked map, the unbiased estimator for $C_{\ell}$ is not straightforward and a large number of methods to obtain it have been suggested in the literature \cite{method1,method4}. In this paper, we adopt the so-called pseudo-$C_{\ell}$ (PCL) estimator method \cite{method4}. Although PCL estimator is a suboptimal one, it can be easily realized in pixel space using fast spherical harmonics transformation, and has been applied to various CMB observations including to WMAP and Planck data. Considering the window function $W(\hat{n})$, the pseudo coefficients $\tilde{a}_{\ell m}$ can be defined as
\begin{equation}
\tilde{a}_{\ell m}=\int \Delta T(\hat{n}) W(\hat{n}) Y_{\ell m}(\hat{n}),
\end{equation}
which is related to $a_{\ell m}$ by
\begin{equation}
\tilde{a}_{\ell m}= \sum_{\ell_{1}m_{1}} a_{\ell_{1}m_{1}} K_{\ell m\ell_{1}m_{1}}.
\end{equation}
The coupling matrix $K$ is given by
\begin{equation}
K_{\ell m\ell_{1}m_{1}}=\sqrt{\frac{(2\ell_1+1)(2l+1)}{4\pi}}\sum_{\ell_2 m_2} (-1)^{m}(2\ell_2+1) w_{\ell_{2}m_{2}}
\begin{pmatrix}
\ell_1 & \ell_2 & \ell \\
0 & 0 & 0
\end{pmatrix}
\begin{pmatrix}
\ell_1 & \ell_2 & \ell \\
m_1 & m_2 & -m
\end{pmatrix},
\end{equation}
and $w_{\ell m}$ are the coefficients of spherical harmonics expansion of the mask $W(\hat{n})$, i.e.,
\begin{equation}
w_{\ell m}=\int W(\hat{n}) Y_{\ell m}^* (\hat{n}) d\hat{n}.
\end{equation}

The pseudo estimator $\tilde{C}_{\ell}$ is defined analogous to (\ref{hat-cl}) in terms of the multipole coefficients $\tilde{a}_{\ell m}$ as
\begin{equation}
\tilde{C}_{\ell}=\frac{1}{2\ell+1} \sum_{m=-\ell}^{\ell} \tilde{a}_{\ell m} \tilde{a}^*_{\ell m}.
\end{equation}
The expectation value of $\tilde{C}_{\ell}$ is
$\langle \tilde{C}_{\ell} \rangle =\sum_{\ell'}C_{\ell'} M_{\ell \ell'}$,
where the coupling matrix is
\begin{equation}
M_{\ell \ell'}=(2\ell'+1)\sum_{\ell_{2}} \frac {2\ell_{2}+1} {4\pi}
\begin{pmatrix}
    \ell'&\ell_2&\ell \\
    0 & 0& 0
    \end{pmatrix}  ^{2}\tilde{w}_{\ell_{2}}
\end{equation}
and ${\tilde w}_{\ell}$ are the following power spectrum,
\begin{equation}
\tilde{w}_{\ell} = \frac{1}{2\ell+1}\sum_{m=-\ell}^{\ell}w_{\ell m}w^{*}_{\ell m}.
\end{equation}
Similarly, the unbiased estimator in the masked sky can be constructed as
$\hat{\mathcal{C}}_{\ell}=\sum_{\ell'} M^{-1}_{\ell \ell'} \tilde{C}_{\ell'}$.
Note that this unbiased estimator $\hat{\mathcal{C}}_{\ell}$ is also rotationally invariant. Actually, the general analyses of the CMB parity asymmetry are always based on estimators of the CMB power spectrum in the masked space \cite{kim2011,planck2013,planck2015}.


In this paper, we focus on the direction dependence of the CMB parity violation. So, the direction dependent estimators are needed in advance. Similar to the discussion above, we can build the direction dependent estimator by excluding the $m=0$ components,
\begin{equation}\label{tilde-dl}
\tilde{D}_{\ell}=\frac{1}{2\ell} \sum_{m=-\ell}^{\ell} \tilde{a}_{\ell m} \tilde{a}^*_{\ell m}(1-\delta_{m0}).
\end{equation}
For each multipole, we have excluded the $m=0$ component, which means that the $z$-direction of the coordinate system is chosen as the preferred direction in the definition. However, the estimators $\tilde{D}_{\ell}$ are not unbiased. The expectation values are given by
$\langle \tilde{D}_{\ell} \rangle =\sum_{\ell'}C_{\ell'} N_{\ell \ell'}$,
where the coupling matrix $N_{\ell \ell'}$ is given by
\begin{equation}
N_{\ell \ell'}=M_{\ell \ell'}-\frac{2\ell' +1}{2\ell} \sum_{\ell_{2} \ell_{2}^{'}m_{1}} \frac {\sqrt{(2\ell_{2} +1)(2\ell_{2}^{'} +1)}} {4\pi}
\begin{pmatrix}
    \ell'&\ell_2&\ell \\
    0 & 0& 0
\end{pmatrix}
\begin{pmatrix}
    \ell'&\ell_2^{'}&\ell \\
    0 & 0& 0
\end{pmatrix}
\begin{pmatrix}
    \ell'&\ell_2&\ell \\
   m_1 & -m_1& 0
\end{pmatrix}
\begin{pmatrix}
    \ell'&\ell_2&\ell \\
    m_1 & -m_1& 0
\end{pmatrix} w_{\ell_2 m_1} w_{\ell_2^{'} m_1}.
\end{equation}
Based on this relation, we can construct the unbiased estimator $\hat{\mathcal{D}}_{\ell}$ as follows,
\begin{equation}
\hat{\mathcal{D}}_{\ell}=\sum_{\ell'} N^{-1}_{\ell \ell'} \tilde{D}_{\ell'}.
\end{equation}
Similar to $\hat{D}_{\ell}$, $\hat{\mathcal{D}}_{\ell}$ are also the coordinate dependent unbiased estimators for the power spectra $C_{\ell}$, and the preferred direction is also the $z$-direction of the corresponding coordinate system.

For any coordinate system, the direction-dependent unbiased estimator $\hat{\mathcal{D}}_{\ell}(\hat{\rm{\bf q}})$ can be built in the same manner with $\hat{\mathcal{D}}_{\ell}$, being, however, the coefficients $\tilde{a}_{\ell m}$ and ${w}_{\ell m}$ replaced by $\tilde{a}_{\ell m}(\hat{\rm{\bf q}})$ and ${w}_{\ell m}(\hat{\rm{\bf q}})$. The direction-dependent statistic for the CMB parity asymmetry can be defined as
\begin{equation}
\mathcal{G}(\ell;\hat{\rm{\bf q}})=\frac{\sum_{\ell'=2}^{\ell}\ell'(\ell'+1)\hat{\mathcal{D}}_{\ell'}(\hat{\rm{\bf q}})\Gamma^{+}_{\ell'}}
{\sum_{\ell'=2}^{\ell}\ell'(\ell'+1)\hat{\mathcal{D}}_{\ell'}(\hat{\rm{\bf q}})\Gamma^{-}_{\ell'}}.
\end{equation}
Since $\hat{\mathcal{D}}_{\ell}$ are the unbiased estimators for the power spectra $C_{\ell}$, the new statistic $\mathcal{G}(\ell;\hat{\rm{\bf q}})$ also indicates the degree of the CMB parity asymmetry and its direction dependence. Comparing with the ideal case for full-sky map and negligible noise, the use of the mask affects the values of statistic $G$ in two aspects: 1) the CMB information is lost in the masked region, and the values of the unbiased estimators for the power spectrum $C_{\ell}$ and their uncertainties might be influenced; 2) the structure and position of the mask may influence the preferred direction of $\mathcal{G}$-maps by the definition of the directional estimator $\tilde{D}_{\ell}$ in Eq. (\ref{tilde-dl}). If the masked region is small, we expect both effects to be negligible, and the results in the masked case should be very close to the ideal case. So, in this paper, we shall only consider the NILC mask suggested by Planck collaboration (see the lower middle panel in Fig. \ref{fig7}).

\begin{figure}[t]
\begin{center}
\includegraphics[width=15cm]{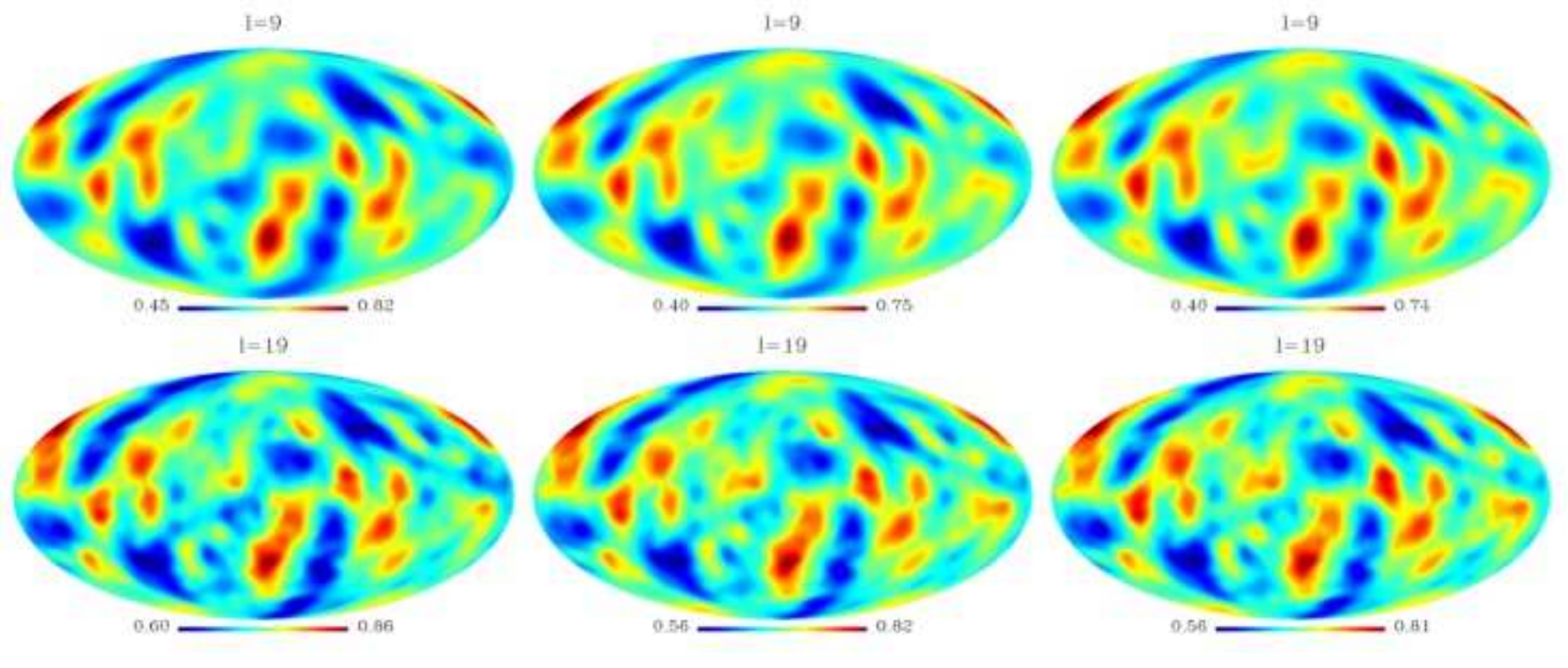}
\end{center}
\caption{The directional statistics $\mathcal{G}(\ell;\hat{\rm{\bf q}})$ for different maximum multipoles $\ell=9$ (upper) and $\ell=19$ (lower) based on the masked Commander (left), NILC (middle) and SMICA (right) maps. Note that, in the figure, we have applied the NILC mask to all the three maps \cite{zhao2015}.}\label{fig8}
\end{figure}

Based on the estimators in the masked maps, we plot the $\mathcal{G}$-maps for different maximum multipole $\ell$ and different CMB maps in Fig. \ref{fig8}, from which we find that the morphological structures of the $\mathcal{G}$-maps are nearly the same for all three input CMB maps. In addition, we also find that they are also quite similar with the results without the mask. In Table \ref{tab6}, as an example, we consider the Planck 2015 NILC map and list the preferred directions for both unmasked and masked cases. All these directions nearly align with each other. This result clearly shows that the preferred directions are independent of the used CMB masks, which stabilize our discovery on the directional properties stored in the CMB parity asymmetry.

\begin{table*}
\caption{The preferred direction $(\theta,\phi)$, and the corresponding $|\cos\alpha|$ and $\Delta_c/\sigma_c$ for $\mathcal{G}(\ell;\hat{\rm{\bf q}})$ based on Planck NILC map, where the different maximum multipole $\ell$ is considered. For each $\ell$ case, the upper values denote the results derived from the full-sky analysis, and the lower values denote those derived from the masked case in which NILC mask is applied \cite{zhao2015}.}
\begin{center}
\label{tab6}
\begin{tabular}{ |c||c |c |c |c |}
    \hline
  &   $~~~~~\theta[^{\circ}]$~~~~~  & ~~~~~$\phi[^{\circ}]$~~~~~  &  ~~~~~$|\cos\alpha|$~~~~~ & ~~~~~$\Delta_c/\sigma_c$~~~~~\\
   \hline
   \hline
   \multirow{3}{*}{$\ell_{\max}=5$} & $45.82$ & $279.73$ & $0.980$ & $3.42$ \\
                                & $45.82$ & $279.73$ & $0.980$ & $3.42$ \\
   \hline
   \multirow{3}{*}{$\ell_{\max}=7$} & $47.41$ & $278.00$ & $0.981$ & $3.39$\\
                                & $48.21$ & $275.06$ & $0.985$ & $3.40$\\
   \hline
   \multirow{3}{*}{$\ell_{\max}=9$} & $48.21$ & $276.47$ & $0.982$ & $3.35$\\
                                & $49.80$ & $272.25$ & $0.985$ & $3.38$\\
   \hline
   \multirow{3}{*}{$\ell_{\max}=11$} & $49.01$ & $277.17$ & $0.979$ & $3.35$\\
                                & $49.80$ & $272.25$ & $0.985$ & $3.38$\\
   \hline
   \multirow{3}{*}{$\ell_{\max}=13$} & $49.01$ & $278.58$ & $0.976$ & $3.34$\\
                                & $49.80$ & $272.25$ & $0.985$ & $3.38$\\
   \hline
   \multirow{3}{*}{$\ell_{\max}=15$} & $49.80$ & $282.10$ & $0.965$ & $3.27$\\
                                & $49.80$ & $272.25$ & $0.985$ & $3.38$\\
   \hline
   \multirow{3}{*}{$\ell_{\max}=17$} & $50.57$ & $284.21$ & $0.957$ & $3.22$\\
                                & $49.80$ & $270.84$ & $0.987$ & $3.39$\\
   \hline
   \multirow{3}{*}{$\ell_{\max}=19$} & $50.57$ & $284.21$ & $0.957$ & $3.22$\\
                                & $49.01$ & $270.14$ & $0.990$ & $3.42$\\
   \hline
   \multirow{3}{*}{$\ell_{\max}=21$} & $50.57$ & $284.21$ & $0.957$ & $3.22$\\
                                & $49.01$ & $270.14$ & $0.990$ & $3.42$\\
   \hline
\end{tabular}
\end{center}
\end{table*}

\section{Comparing with the CMB low multipoles}

\subsection{CMB kinematic dipole}

\begin{figure*}[t]
\begin{center}
\includegraphics[width=10cm]{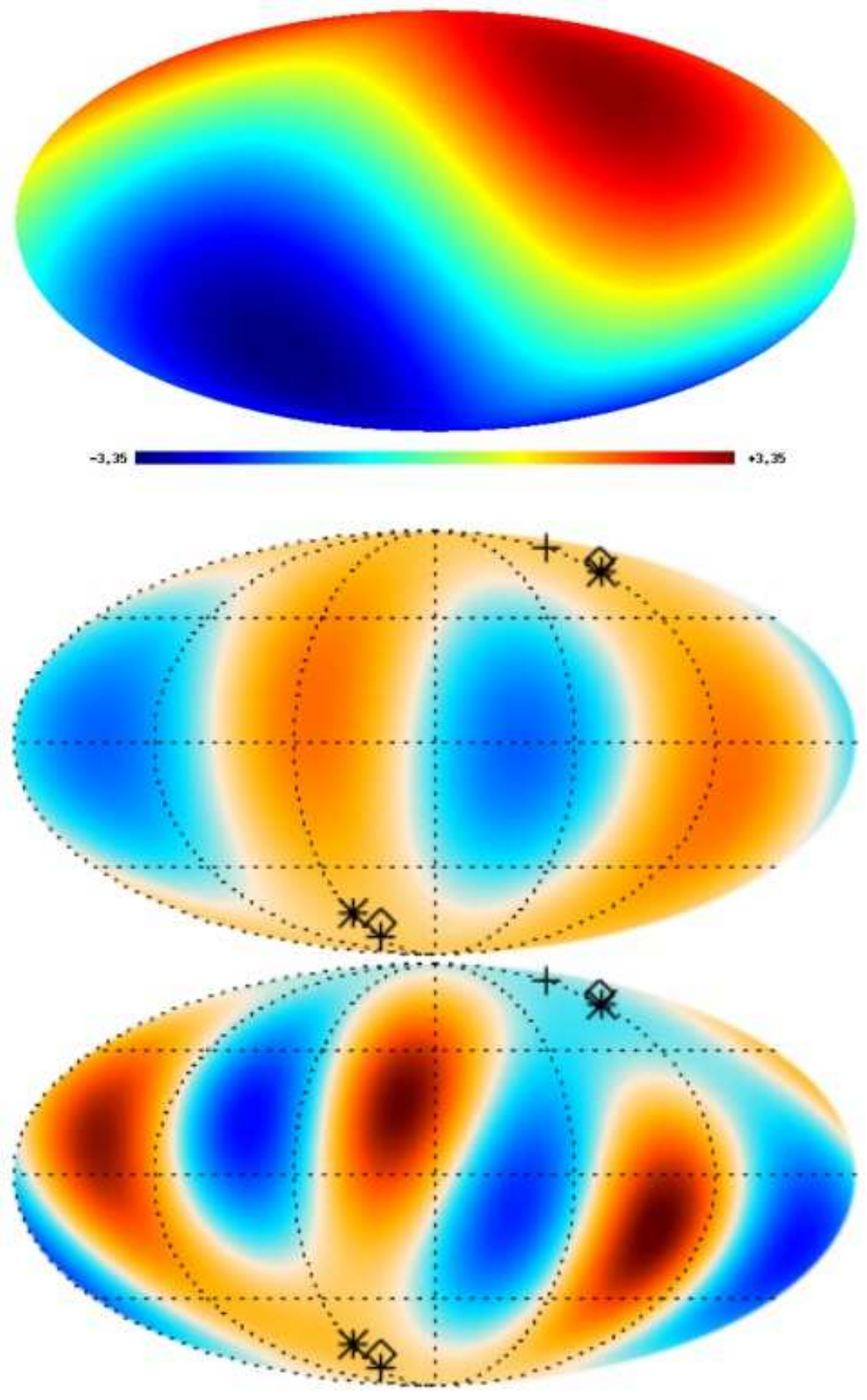}
\end{center}
\caption{The upper panel is the CMB kinematic dipole with in mK. Middle panel is the quadrupole (temperature range $\pm$ 35 $\mu$K) derived from wiener-filtered SMICA CMB sky, and lower panel is the derived octopole (temperature range of $\pm$ 35 $\mu$K).  The plus and the star symbols indicate the axes of the quadrupole and octopole, respectively, around which the angular momentum dispersion is maximized. The diamond symbols correspond to the quadrupole axes after correcting for the kinematic quadrupole \cite{1303.5083}.}\label{fig9}
\end{figure*}

As well known, the CMB has an extremely uniform temperature of 2.725 Kelvin, which is a left over from the period of recombination. The lowest anisotropy is the dipole component with an amplitude of $3.35$ mK from the Doppler shift of the background radiation \cite{dipole} caused by the 
peculiar velocity of solar system at about $370$ km/sec relative to the comoving cosmic rest frame the dipole anisotropy defines a peculiar axis in the Universe, which is at $(\theta=42^{\circ}, \phi=264^{\circ})$ in Galactic coordinate system \cite{dipole} (see left panel of Fig. \ref{fig9}) \footnote{Note that, an alternative explanation for the CMB dipole is also discussed in \cite{alternative-dipole}}.

In Figs. \ref{fig4}, \ref{fig5} and \ref{fig7}, we compare the preferred directions $\hat{\bf
q}$ of parity asymmetry with the CMB kinematic dipole and find that
they are very close to each other.
Especially, all these directions are close to the ecliptic
plane. To quantify it, we define the quantity $\alpha$, which is
the angle between $\hat{\bf q}$ and the CMB kinematic dipole
direction at $(\theta=42^{\circ},\phi=264^{\circ})$ \cite{dipole}.
In Tables \ref{tab2}, \ref{tab3} and \ref{tab6}, we list the values of $|\cos\alpha|$ in the corresponding cases, and find that all of them are very close to $1$. For instance, by using the most recent Planck NILC map (see Table \ref{tab6}), in both masked case and unmasked case, we get that $|\cos\alpha|>0.98$ holds for all maximum multipole cases, which means that the angles between the preferred direction $\hat{\rm{\bf q}}$ and the CMB dipole direction are all smaller than $11.5^{\circ}$. In Tables \ref{tab4} and \ref{tab5}, we also list the corresponding values of $|\cos\alpha|$ derived from the other CMB maps based on six different statistics, and find similar results. So, we get a stable conclusion: \emph{The preferred direction in the CMB parity asymmetry strongly aligns with the CMB kinematic dipole, which is independent of the CMB maps, the directional statistics or the used mask.} This coincidence strongly indicates that the anomaly on the CMB parity asymmetry, as well as its directional properties, may be related to the CMB kinematic dipole component.

\begin{table*}
\caption{The values of $|\cos\alpha|$ and $\Delta_c/\sigma_c$ for the
statistics $G_i(\ell;\hat{\bf q})$ based on Planck 2012 year NILC and SEVEM
data. Similar to Table \ref{tab2}, in each box, the upper one is the result
for the statistic with $i=1$, the middle one is that for $i=2$, and the
lower one is that for $i=3$.}
\begin{center}
\label{tab4}
\begin{tabular}{ |c||c |c| |c |c|  }
    \hline
     & ~~~$|\cos\alpha|$ for NILC~~~ & ~~~$\Delta_c/\sigma_c$ for NILC~~~ & ~~~$|\cos\alpha|$ for SEVEM~~~ & ~~~$\Delta_c/\sigma_c$ for SEVEM~~~   \\
   \hline
   \hline
   \multirow{3}{*}{$\ell_{\max}=3$} &    0.3259  &  0.88    & 0.2656 &    0.66      \\
   &    0.3259  &  0.88    & 0.2656 &    0.66      \\
   &    0.3259  &  0.88    & 0.2656 &    0.66      \\
   \hline
   \multirow{3}{*}{$\ell_{\max}=5$} &   0.9758  &   3.38   & 0.9802 &     3.42   \\
   &   0.9748  &   3.36   & 0.9802 &     3.42   \\
   &   0.9748  &   3.36   & 0.9802 &     3.42    \\
   \hline
   \multirow{3}{*}{$\ell_{\max}=7$} &   0.9822  &   3.37   & 0.9840 &     3.41    \\
   &   0.9769  &   3.36   & 0.9812 &     3.39    \\
   &   0.9812  &   3.33   & 0.9861 &     3.37    \\
   \hline
   \multirow{3}{*}{$\ell_{\max}=11$} &   0.8600  &   3.18   & 0.8600 &     3.18    \\
   &   0.9697  &   3.29   & 0.9766 &     3.34   \\
   &   0.8520  &   3.15   & 0.8600 &     3.18     \\
   \hline
   \multirow{3}{*}{$\ell_{\max}=21$} &   0.8932  &   3.25   & 0.9529 &     3.19     \\
   &   0.9575  &   3.22   & 0.9575 &     3.22    \\
   &   0.8522  &   3.13   & 0.5295 &     2.79   \\
   \hline
\end{tabular}
\end{center}
\end{table*}

\begin{table*}
\caption{The values of $|\cos\alpha|$ and $\Delta_c/\sigma_c$ for the statistics $G_i(\ell;\hat{\bf q})$ based on Planck 2013 year NILC and SEVEM data. Similar to Table \ref{tab3},
in each box the upper one is the result for the statistic with $i=4$, the middle one is that for $i=5$ and the lower one that is for $i=6$.}
\begin{center}
\label{tab5}
\begin{tabular}{ |c||c |c| c |c|  }
    \hline
     & ~~~$|\cos\alpha|$ for NILC~~~ & ~~~$\Delta_c/\sigma_c$ for NILC~~~ & ~~~$|\cos\alpha|$ for SEVEM~~~ & ~~~$\Delta_c/\sigma_c$ for SEVEM~~~   \\
   \hline
   \hline
   \multirow{3}{*}{$\ell_{\max}=3$} &    0.3094  &  0.78    & 0.2650 &    0.64      \\
   &    0.3094  &  0.78    & 0.2650 &    0.64      \\
   &    0.3094  &  0.78    & 0.2650 &    0.64      \\
   \hline
   \multirow{3}{*}{$\ell_{\max}=5$} &   0.8705  &   3.06   & 0.9040 &     3.16   \\
   &   0.8705  &   3.06   & 0.8963 &     3.13   \\
   &   0.8617  &   3.03   & 0.8963 &     3.13    \\
   \hline
   \multirow{3}{*}{$\ell_{\max}=7$} &   0.9838  &   3.37   & 0.9585 &     3.14    \\
   &   0.9852  &   3.43   & 0.9710 &     3.26    \\
   &   0.9782  &   3.38   & 0.9693 &     3.28    \\
   \hline
   \multirow{3}{*}{$\ell_{\max}=9$} &   0.9097  &   3.73   & 0.9351 &     3.74    \\
   &   0.9450  &   3.77   & 0.9713 &     3.77   \\
   &   0.9290  &   3.75   & 0.9529 &     3.77   \\
   \hline
\end{tabular}
\end{center}
\end{table*}

\subsection{The CMB quadrupole and octopole}

The lowest cosmological anisotropic modes of the CMB fluctuations are the quadrupole and octopole. By defining the directional statistic and applying it to the first year WMAP data, Tegmark et al. found that both CMB quadrupole and octopole point to preferred directions \cite{alignment0}. It was also found that their orientation are strongly aligned with each other. Slightly later, by alternative methods, the alignment was confirmed by several groups \cite{alignment}. In addition, \cite{alignment2}  found that the alignment of the CMB low multipole can extend to $\ell=5$, which is a significant violation of the random Gaussian assumption of the primordial fluctuations. Recently, Planck collaboration applied similar analysis to the Planck data \cite{1303.5083}. For each multipole $\ell$, they determine the orientation of the multipoles by finding the axis $\hat{n}$ around the maximized angular momentum dispersion.
\begin{equation}
\sum_{m} m^2 |a_{\ell m}(\hat{n})|^2
\end{equation}
Using the Planck 2013 SMICA map and considering the U73 mask, Planck collaboration constructed the quadrupole and the octopole, determining their preferred directions: $(\theta=13.4^{\circ}, \phi=238.5^{\circ})$ and $(\theta=25.7^{\circ}, \phi=239.0^{\circ})$, respectively. The angular difference between these directions is of only $12.3^{\circ}$, and the significance of the alignment is of $96.8\%$. In Fig. \ref{fig9} (middle and lower panels), we plot the constructed multipole components, and the corresponding preferred directions. The alignment between them is clearly shown. Using different maps, or considering the correction for the kinematic quadrupole (denoted as KQ corrected), the preferred directions of the quadrupole and the octopole are slightly different. They are listed in Table \ref{tab8}. However, in each case, the alignment holds at quite high confidence level.

\begin{table}
\caption{Preferred directions of CMB quadrupole and octopole in the Planck 2013 year data \cite{1303.5083}.}
\begin{center}
\label{tab8}
\begin{tabular}{ |c|c|c|c| }
   \hline
   CMB map & $(\theta,\phi)$ for quadrupole [degree] &$(\theta,\phi)$ for octopole [degree]&angle between them [degree]\\
   \hline
   C-R  &    $(29.7,228.2)$ & $(24.0,246.1)$ & 9.80        \\
   NILC & $(12.7,241.3)$ & $(25.8,241.7)$ & 13.1 \\
   SEVEM & $(16.2,242.4)$ & $(25.2,245.6)$ & 9.08\\
   SMICA & $(13.4,238.5)$ & $(25.7,239.0)$ & 12.3 \\
   NILC, KQ corrected & $(20.3,225.6)$ & $(25.8,241.7)$ & 8.35\\
   SEVEM, KQ corrected & $(21.7,228.3)$ & $(25.2,245.6)$ & 7.69 \\
   SMICA, KQ corrected & $(20.8,224.2)$ & $(25.7,239.0)$ & 7.63 \\
      \hline
\end{tabular}
\end{center}
\end{table}


In this paper, we shall investigate whether or not
the alignment of quadrupole and octopole connects with the parity
asymmetry. Following Ref. \cite{anto}, it is
straightforward to evaluate the mean value of the inner product
between all the pairs of unit vectors corresponding to the
following four directions: The preferred directions of quadrupole,
octopole, parity asymmetry and the direction of the CMB kinematic
dipole. So, we define the quantity,
 \be\label{ctheta}
 \langle |\cos\theta_{ij}|\rangle = \sum_{i,j=1,~j\neq i}^{N} \frac{|\hat{r}_i\cdot \hat{r}_j|}{N(N-1)},
 \ee
where $N$ is the number of the directions that will be
investigated. First, we shall study the case in the absence of the
parity asymmetry. The alignment between these three axes was also
reported in WMAP data \cite{schwarz}. In this case, we have $N=3$
and $\langle |\cos\theta_{ij}|\rangle=0.9242$ for the real data.
{To evaluate the significance of the alignment, we pixelize the two-dimensional sphere in the HEALPix format with the resolution parameter $N_{\rm side}=256$, which corresponds to the total pixel number $N_{\rm pix}=12\times N_{\rm side}^2$. Then, we randomly generate $10^5$ realizations. For each realization, all three directions are randomly and independently picked on the sky using a uniform distribution between 0 and $N_{\rm pix}-1$,} and the corresponding value $\langle |\cos\theta_{ij}|\rangle$ is calculated directly. Considering all the random samples, we obtain that $\langle |\cos\theta_{ij}| \rangle = 0.500 \pm 0.167$. {To quantify the significant level of the deviation from the random distribution, we define the
$\Delta_c/\sigma_c$, where $\Delta_c$ is the difference between the observed value of $\langle |\cos\theta_{ij}|\rangle$ and the mean value of the simulations, and $\sigma_c$ is the corresponding standard deviation of the simulations. Considering the observed result $\langle |\cos\theta_{ij}|\rangle=0.9242$, and the simulated value $\sigma_c=0.167$, we obtain that $\Delta_c/\sigma_c=2.54$, which indicates that the alignment of these three directions is around $2.5\sigma$ confidence level.}

\begin{figure*}[t]
\begin{center}
\includegraphics[width=13cm]{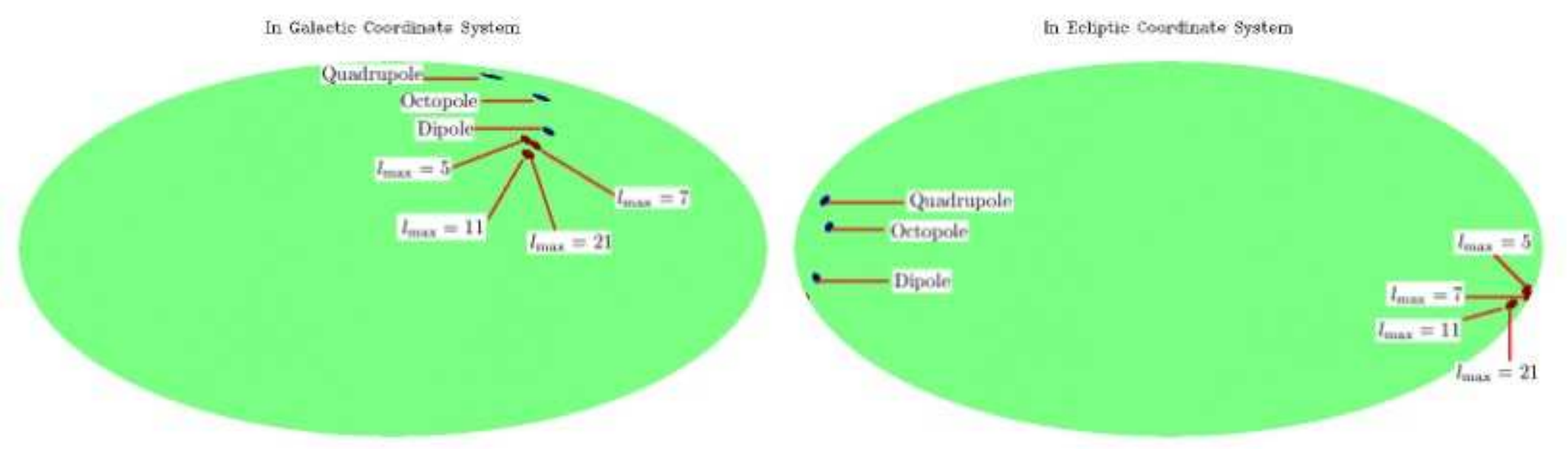}
\end{center}
\caption{The preferred directions of the SMICA-based statistics
$G_1(\ell;\hat{\bf q})$ in the Galactic coordinate
system (left) and in the ecliptic coordinate system (right). In both
panels, we have compared them with the CMB kinematic dipole
direction and the preferred directions of the CMB quadrupole and
octopole.}\label{fig10}
\end{figure*}

Now, let us take into account the preferred direction of the CMB
parity asymmetry. For the $10^5$ random realizations of four
random points on the sphere, by a similar analysis, we get
 \be\label{theta4}
 \langle |\cos\theta_{ij}| \rangle = 0.500 \pm 0.118.
 \ee
As anticipated, compared with the case of $N=3$, the mean value
stays the same, while the standard deviation $\sigma_c$ significantly decreases. Now, we
calculate the real value of the quantity $\langle
|\cos\theta_{ij}| \rangle$. For the preferred direction of parity
asymmetry, we consider the results of all the six statistics
$G_i(\ell;\hat{\bf q})$ defined in this paper, and list the
corresponding values in Tables \ref{tab1}-\ref{tab5}. For every
case with $\ell>3$, we find that $\langle |\cos\theta_{ij}| \rangle$
is close to $0.9$, and the corresponding significance of the
alignment between these directions increases to $\Delta_c/\sigma_c \gtrsim 3$.
In Fig. \ref{fig10}, we plot all these preferred directions in the Galactic coordinate system (left) and ecliptic coordinate system (right), and clearly present the alignment of them.

We therefore conclude that \emph{the preferred direction of the CMB
parity asymmetry is not only very close to the CMB kinematic
dipole, but also close to the preferred axes of the CMB quadrupole
and octopole, which is nearly independent of the choice of the
parity statistic, and the used CMB map or mask}.  Since the quadrupole and the octopole also relate to other CMB anomalies, including the low quadrupole problem and the lack of the large-scale correlation, the coincidence between the preferred directions in the CMB low multipoles also hints: \emph{The CMB parity asymmetry is not an isolated anomaly, it may have the common origin with all these CMB anomalies.}

\section{Preferred axes in other large-scale observations}

In addition to the directional problem in the CMB low multipoles, in other cosmological observations,  similar preferred axes were also reported in the literature. In particular, several axes are announced to align with the CMB kinematic dipole and in this section, we briefly list them as below.

\subsection{Alignment of quasar polarization vectors}
Quasars are the most energetic and distant members of a class of objects called active galactic nuclei.  Since quasars show very high redshifts, they can be treated as the trackers to study the matter distributions of the Universe. Based on the sample of 170 polarized quasars, in 1998, Hutsemekers studied the distribution of the quasar polarization vectors \cite{quasar}. In general, we naturally expect the quasar polarization angles to be randomly distributed between $0^{\circ}$ and $180^{\circ}$. However, by applying two different statistical tests, the author found that the optical polarization vectors of quasars are not randomly distributed over the sky but are coherently oriented on very large spatial scales. The instrumental bias and the contamination by interstellar polarization in our Galaxy are unlikely to be responsible for these features. Lately, different groups and different analyses \cite{quasar2} also derived similar results, i.e. the quasar polarization vectors are not randomly oriented over the sky with a probability often in excess of $99.9\%$. The alignment effect seems to be prominent along a particular axis in the direction $(\theta=69^{\circ},\phi=267^{\circ})$ in the Galactic coordinate system (see Table \ref{tab7}), which seems coincident with the axis defined by the CMB kinematic dipole.

\subsection{Large-scale velocity flows}
Since the discovery of the cosmic microwave background radiation, people realized that one absolute reference system is defined, in which the CMB photons have no velocity flows. However, the regular matters, including baryons and dark matter, may have bulk flow relative to the CMB reference coordinate. The peculiar velocity field of the galaxies provides an important and robust way to understand the matter distribution and motion of the universe, which can be used to detect possible unobservable sources of gravitational fields. Recently, the peculiar velocity field of the nearby universe has been explored extensively using variant velocity tracers \cite{chen,velocity}. It was found that the flow amplitudes and directions strongly depends on the scales which have been focused on. In particular, on the intermediate scales from $20h^{-1}$Mpc to about $100h^{-1}$Mpc, there is evidence of a bulk flow with amplitude $416\pm 78$ ${\rm km~s}^{-1}$ at the direction of $(\theta=84^{\circ}\pm 6^{\circ},\phi=282^{\circ}\pm11^{\circ})$ \cite{velocity}. In the standard $\Lambda$CDM model, this flows can be generated by the primordial fluctuations with long-wavelength modes. However, the predictive amplitude at this scale is only approximately $100$ ${\rm km~s}^{-1}$, and the probability that a flow of magnitude larger than 400 ${\rm km~s}^{-1}$ is realized is less than $0.5\%$. This is the so-called dark flow problem for the galaxies. However, we should also mention that this problem is still in debate. Some other authors claimed the normal bulk velocities on this scale using separate velocity catalogs \cite{velocity2}.

\subsection{Handedness of spiral galaxies}
Another directional anomaly was found in the spiral galaxies. Using 15158 spiral galaxies with redshifts $z<0.085$ from the Sloan Digital Sky Surveys,  in \cite{spiral} the author studied the left-handed and right-handed spirals of these galaxies and their distribution in the sky. Surprisingly enough, it was found that the handedness distribution of these spiral galaxies are not random. A significant dipole component with the amplitude $A=-0.0408\pm 0.011$ was discovered at more than $99.9\%$ confidence level. The observed spin correlation extends out to separations $\sim 210{\rm Mpc/h}$. Defining the axis of the asymmetry along to the direction of $L$ ($\circlearrowright$) excess, the preferred direction lies near $(\theta=158.5^{\circ},\phi=232^{\circ})$ (or the equivalent direction $(\theta=21.5^{\circ},\phi=52^{\circ})$) in Galactic coordinates, which seems to align with the preferred axis of spin vector in our galaxy. This distribution asymmetry and the alignment indicate a parity violation in the overall Universe.

\subsection{Anisotropy of the cosmic acceleration}
Soon after the discovery of the accelerating cosmic expansion from the observation of Type Ia supernova, a number of authors have investigated the anisotropies of this cosmic acceleration. In particular, several groups have applied the hemisphere comparison method to study the anisotropy of cosmic acceleration \cite{acceleration}. By comparing the supernova data and the corresponding cosmic accelerations on several pairs of opposite hemispheres, a statistically significant preferred axis has been reported in literature. Recently, this study is extended by Javanmard et al. in \cite{acc2}. Using the new leased Union 2.1 compilation, the authors singled out the most discrepant direction with the respect to the all-sky data, and built maps with three different angular resolutions to test the isotropy of the magnitude-redshift of supernovae. It was found that the null hypothesis should be rejected at $95\%$-$99\%$ confidence level, and the strongly deviations in the Union2.1 sample occurs at $(\theta=23.4^{\circ},\phi=247.5^{\circ})$, which closely align with the CMB kinematic dipole.

However, we should emphasize that this conclusion seems not to be stable now. By using different data samples, and/or different analysis methods, people have derived quite different results in the recent literature \cite{debate1,debate2,debate3}. For example, in our paper \cite{debate3}, we divided the Unions dataset into 12 subsets according to their positions in the Galactic coordinate system. In each region, we derived the deceleration parameter $q_0$ as the diagnostic to quantify the anisotropy level in the corresponding direction.  The dipole component was also found in the $q_0$-map at more $95\%$ confidence level. The direction of the best-fit dipole is $(\theta=108.8^{\circ},\phi=180.7^{\circ})$, which is quite different from the direction found in \cite{acceleration} and \cite{acc2}. So, further analysis and much better data are needed to clarify this debate in the future.

\subsection{Anisotropic distribution of fine-structure constant}

On the sightline from us to the quasar, there are numerous gas clouds that absorbs the quasar radiation. Spectroscopic observations can reveal the absorption spectrum of the atoms, ions and molecules of the intervening clouds. These absorption spectra have been used to search for the space-time variation of the fine-structure constant $\alpha$ in cosmology \cite{fine}. Using more than one-decade data of Keck telescope and Very Large Telescope, Webb et al. reported evolution of $\alpha$ in the Universe, which significantly violate the prediction of the standard model of particle physics. The evolution tendencies of $\alpha$ are also different at different regions in the sky. It was shown that they well fit the angular dipole model. The dipole axis was found to point in the direction ($\theta=104^{\circ},\phi=331^{\circ}$) and the dipole amplitude was found to be $A=(0.97\pm0.21)\times10^{-5}$, which excludes the isotropic hypothesis at more than $4\sigma$ confidence level \cite{fine}. The discovered preferred axis in $\alpha$ evolution is claimed to strongly coincide with the other axes, in particular the preferred axis in the anisotropy of the cosmic acceleration \cite{acceleration}. So, it was suggested that at least these two anomalies have a common origin \cite{fine-common}: the anisotropy of the cosmic acceleration and the fine-structure constant anomaly. However, we also need to mention that there are still some debates regarding this conclusion \cite{debate4,debate5}.

\begin{table}
\caption{Preferred directions in various large-scale observations}
\begin{center}
\label{tab7}
\begin{tabular}{ |c|c|c|  }
   \hline
   observations  & ~~~~~~~~~~~$\theta$ [degree]~~~~~~~~~~~ & ~~~~~~~~~~~$\phi$ [degree]~~~~~~~~~~~ \\
   \hline
   CMB kinematic dipole  &    42 & 264       \\
   CMB quadrupole  &    13.4 & 238.5       \\
   CMB octopole  &    25.7 & 239.0       \\
  CMB parity asymmetry & 45.82 & 279.73 \\
  Polarization of QSOs & 69 & 267 \\
  Large-scale velocity flows & 84 & 282 \\
  Handedness of spiral galaxies & 158.5 & 232 \\
  Anisotropy of cosmic acceleration & 23.4 & 247.5 \\
  Distribution of fine-structure constant & 104 & 331 \\
      \hline
\end{tabular}
\end{center}
\end{table}

\subsection{Dipole observations with radio galaxy catalog}

The peculiar velocity of the Solar System relative to the frame of distant radio sources were also studied by \cite{radio_sources} using the data of the NRAO VLA Sky Survey (NVSS). In these papers, even though, the direction of the dipole is in agreement with the CMB previous results, its magnitude was found to be much larger than the CMB expectations for the number counts or sky brightness observables. Presently, the NVSS dipole is in disagreement with the CMB predicted velocity dipole at 2.7$\sigma$ C.L.  In addition, if the linearly polarised flux density is considered instead, the level of anisotropy increases in comparison to pure number counts or sky brightness. The authors suggest that if the difference between the two dipoles is confirmed, it would imply an intrinsically anisotropic universe, with the anisotropy changing with the epoch. Moreover, the axis of anisotropy would point roughly towards the CMB dipole direction. However, recent results of the NVSS dipole leaded to the conclusion that a large bias factor for radio galaxies at low redshift could explain the NVSS dipole signal. It is important to point out that several other explanations for the NVSS dipole are still considered in literature, and it remains a puzzle.

\section{Possible interpretations}

From the discussions above, we know the preferred axes have been found in a number of large-scale observations. And also, at least some of them seem to align with each other. These coincidences might imply the same origin for these anomalies and their interpretations can be divided into two different kinds: In the first kind of explanations, the anomalies have a cosmological origin. Since the existence of the preferred axis in the universe is a significant violation of the cosmological principle, if these anomalies are due to some cosmological effect, one has to consider an alternative model to replace the standard $\Lambda$CDM model. In the second kind of explanations, the standard cosmological model is correct, while the current observations or data analysis of the large-scale data are still inaccurate. These anomalies may be caused by some unsolved systematics or contaminations. In the following discussion, we will briefly introduce these interpretations, respectively.

\subsection{Non-trivial topology of the universe}

As mentioned before, the standard cosmological model is based on two assumptions:  One is that Einstein's general relativity correctly describes gravity, the other assumes the universe as homogeneous and isotropic on large scales. If we believe that the anomalies have a cosmological origin, at least one of these two assumptions will be broken. In \cite{1303.5083,1303.5086}, Planck collaboration investigated several non-trivial topological models in the universes with locally flat, hyperbolic and spherical geometries. Unfortunately, no evidence has been found in the observed data.

Another possibility relies on the Bianchi models. The Bianchi classification provides a complete characterization of all the known homogeneous but anisotropic exact solution to general relativity \cite{ellis}. So, in general, Bianchi models can provide preferred directions in the universe (see for instance \cite{ellip}). Among them, Bianchi VII$_{h}$ models describe a universe with overall rotation, parameterized by an angular velocity and a three-dimensional rate of shear (including the case with zero angular velocity and non-zero shear \cite{john}). In this model, a free parameter is defined to describe the comoving length-scale over with the principal axes of shear and rotation change orientation, and many authors have studied the imprints of these models in the CMB data \cite{jaffe}. In \cite{1303.5086}, it was found that Planck data do provide evidence supporting a phenomenological Bianchi VII$_{h}$ component. However, a physical, anisotropic Bianchi universe is not supported by the data.

\subsection{Alternative gravitational theories}

In order to explain the bulk flow, in \cite{1009.1509} the authors considered that the universe is influenced by large-scale ``wind", and the cosmic matter is drifted by this ``wind". The velocity of the ``wind" takes account for the observed peculiar velocity. When the ``wind" has a privileged direction, the cosmic matter drifts towards the same direction. Actually, the ``wind" picture refers to the Zermelo navigation problem, which is described by the Finsler geometry \cite{bao2000}. Finsler geometry is a natural framework for describing an anisotropic spacetime. The study of the metric is the fourth root of a quartic differential form. Chang et al. found that if Riemann geometry is replaced by Finsler geometry, and Einstein's general relativity is replaced by the general relativity based on Finsler geometry, the observed bulk flow can be naturally explained. In addition, it was found that a matter dominated navigation cosmological model could account for the accelerating expansion of the universe \cite{li2012}, in which the anisotropy of the cosmic acceleration is also possible to be explained \cite{1009.1509}.

Another theoretical explanation of the observed preferred direction is suggested by Yan et al. \cite{deSitter}, which is motivated by the fact that the cosmological constant $\Lambda$ is nonzero. So, the metric of the local inertial reference system in the standard model of cosmology is the Beltrami metric instead of the Minkowski one, and the basic spacetime symmetry has to be from de Sitter's group. To avoid ambiguities caused by the inertial forces, quantum mechanics for spectra in atoms are defined in inertial coordinate systems. The corresponding special relativity is de Sitter special relativity, which is proposed in the literature. In this model, the Minkowski point does exist in the universe, where the Beltrami metric returns to the Minkowski one, and the physics at this point returns to the Einstein's special relativity. The Minkowski point naturally defines a preferred point in the universe, and the direction pointing is the preferred direction. When extending this theory to the general relativity, i.e. de Sitter general relativity, the authors found that the evolution of fine-structure constant $\alpha$, and its dipole structure can be well explained \cite{deSitter}.

\subsection{Particular fluctuation modes or dark energy models}

Actually, the explanation for the directional anomalies with a minimum cost is to consider the possible anisotropic matter component and/or superhorizon fluctuation modes in the universe. In the present stage of universe, the dark energy is the dominant component. If this component is anisotropic, the observed universe may display special directions. In the general scenario, the dark energy is described by the cosmological constant $\Lambda$ or some scalar field. However, it is also possible that the dark energy can be described by a vector field. For instance, in \cite{ym}, we suggested to use the quantum Yang-Mills condensate to describe the dark energy and promote the cosmic acceleration. For the vector fields, the spatial distribution is always anisotropic, which could easily lead to a preferred axis in the large scales. Besides, in \cite{fine-common}, some other special dark energy models have been proposed to explain the anisotropic of cosmic acceleration or the anisotropic distribution of the fine-structure constant, providing another possibility to explain the large-scale anomalies. Among them, we can cite anisotropic dark energy models, inhomogeneous dark energy models, topological quintessence, and spherical dark energy overdensity.

For the large-scale CMB anomalies, the mechanism of Grishchuk-Zeldovich effect is a possible explanation \cite{gz-effect}. It is the contribution to the CMB temperature anisotropy from an extremely large-scale adiabatic density perturbation, considering the standard hypothesis that this perturbation is a typical realization of an homogeneous Gaussian random field. Assuming that one particular large-scale mode of fluctuation, generated in the early inflationary stage, dominates the perturbations of the largest scales in the current stage of universe, then this mode necessarily indicates a preferred direction in the universe. It could also significantly influence the CMB quadrupole and octopole, which might answer the alignment problem of CMB low multipoles.

\subsection{Unsolved systematical errors or contaminations}
On the contrary, some people believe that the standard model of cosmology, based on the cosmological principle and general relativity, is an accurate model to describe the current universe. Considering that the observed large-scale directional anomalies have a non-cosmological origin, being therefore caused by some unsolved systematical errors, calibration errors or contaminations, we list some possibilities.

One possible explanation is related to the contaminations generated by the collective emission of Kuiper Belt objects and other minor bodies in the solar system where the kinematic dipole of CMB is located. Since the emission of the Kuiper Belt objects is nearly independent of the frequency, this contamination is very hard to remove from the CMB data analysis. In Maris et al. \cite{maris} and Hansen et al. \cite{hansen}, it was discussed that this foreground residual could leave significant imprints in the CMB low multipoles and possibly explain the CMB parity asymmetry, as well as the alignment of the CMB quadrupole and octopole.

Another explanation may relate to a deviation measured in the CMB kinematic dipole, which could be due to a measurement error in the dipole direction, a problem in the antenna pointing direction, sidelobe pickup contamination, and so on. In \cite{liu2011}, it was found that this kinematic dipole deviation could generate the artificial CMB anisotropies in the low multipoles. If this is true, these artificial components may account for the CMB large-scale anomalies.

It is also possible that the preferred direction is caused by the tidal field originated from the anisotropy of our local halo. In \cite{wang}, the authors found that the tidal field tends to preferentially align with the orientation and spatial distribution of galaxies, which may also generate some unsolved kinematic or higher order effects, and influence the cosmological observations \cite{wang2}.

\section{Conclusions and discussions}

The standard $\Lambda$CDM model has a great success in explaining the observations of the CMB temperature anisotropies, as well as the galaxies distribution and motion. The standard model of cosmology is based on the assumptions: the validity of Einstein's general relativity, and the cosmological principle. This model can explain most large-scale observations with unprecedented accuracy. However, several directional anomalies have been reported in various observations: the polarization distribution of the quasars, the velocity flow, the handedness of the spiral galaxies, the anisotropy of the cosmic acceleration, the anisotropic evolution of fine-structure constant, including anomalies in the CMB low multipoles, such as the CMB parity asymmetry. Although the confidence level for each individual anomaly is not too high, the directional alignment of all these anomalies is quite significant, which strongly suggests a common origin of these anomalies.

If these anomalies are due to cosmological effects , e.g. the alternative theory of gravity or geometry, the non-trivial topology of the universe, the anisotropic dark energy or the particular large-scale fluctuation modes, they indicate the violation of the cosmological principle. So, one should consider to build a new cosmological model to explain the large-scale data. However, if these directional anomalies arise from non-cosmological reasons, e.g. the unsolved systematical errors or contaminations, we should carefully treat the current data, and exclude the errors in the future analysis to avoid the misleading explanations of the data.

In order to distinguish these two kinds of explanations, we compare the preferred directions in large-scale observations and the CMB kinematic dipole, and found a strong alignment between them. As well known, the CMB dipole is caused by the motion of the Solar System in the universe, which is a purely kinematic effect. The alignment of CMB dipole and the other preferred direction strongly suggest a non-cosmological origin of the large-scale anomalies, which should be caused by some CMB dipole-related systematics or contamination. In future cosmological observations, we suggest to further study these possible errors, and subtract them from the observed data. In addition, we expect that the future measurements on the CMB polarization fields, the cosmic weak lensing, or the distribution of 21-cm line can help us to solve the puzzles.

\section*{Acknowledgements}
We acknowledge the use of the Legacy Archive for Microwave Background Data Analysis (LAMBDA) and Planck Legacy Archive (PLA). Our data analysis made the use of HEALPix
\cite{healpix}. This work is supported by Project 973 under Grant No. 2012CB821804, by NSFC No. 11173021, 11322324, 11421303 and project of KIP and CAS.

\end{document}